\documentclass[12pt]{article}
\usepackage{amsmath} 
\usepackage{multirow} 
\usepackage{amsfonts}
\usepackage{amssymb,graphics,psfrag}
\usepackage{array,epsfig,multirow,stmaryrd,graphicx,mathtools}
\usepackage{comment}
\def\hybrid{\topmargin -20pt    \oddsidemargin 0pt
        \headheight 0pt \headsep 0pt
        \textwidth 6.25in       
        \textheight 9 in       
        \marginparwidth .875in
        \parskip 5pt plus 1pt 
          \jot = 1.5ex
   }
\hybrid
\numberwithin{equation}{section}
\numberwithin{table}{section}

\newcommand{\beq}{\begin{equation}}
\newcommand{\eeq}{\end{equation}}
\newcommand{\be}{\begin{equation}}
\newcommand{\ee}{\end{equation}}
\newcommand{\bea}{\begin{eqnarray}}
\newcommand{\eea}{\end{eqnarray}}   
\newcommand{\ben}{\begin{eqnarray*}}
\newcommand{\een}{\end{eqnarray*}}                  
\newcommand{\ba}{\begin{aligned}}
\newcommand{\ea}{\end{aligned}}
\newcommand{\bt}{\begin{tabular}}
\newcommand{\et}{\end{tabular}}
\newcommand{\bc}{\begin{center}}
\newcommand{\ec}{\end{center}}

\newcommand{\cO}{\mathcal{O}}
\newcommand{\cT}{\mathcal{T}}

\newcommand{\cC}{\mathcal{C}}
\newcommand{\cD}{\mathcal{D}}
\newcommand{\cL}{\mathcal{L}}
\newcommand{\cS}{\mathcal{S}}
\newcommand{\cK}{\mathcal{K}}
\newcommand{\cN}{\mathcal{N}}
\newcommand{\cW}{\mathcal{W}}
\newcommand{\cG}{\mathcal{G}}
\newcommand{\cA}{\mathcal{A}}

\newcommand{\cB}{\mathcal{B}}
\newcommand{\cF}{\mathcal{F}}

\newcommand{\cV}{\mathcal{V}}

\newcommand{\cM}{\mathcal M}

\newcommand{\I}{\text{Im}}
\newcommand{\R}{\text{Re}}

\newcommand{\bi}{{\bar \imath}}

\newcommand{\bj}{{\bar\jmath}}

\newcommand{\bbZ}{\mathbb{Z}}
\newcommand{\bbR}{\mathbb{R}}
\newcommand{\bbC}{\mathbb{C}}
\newcommand{\bbP}{\mathbb{P}}

\newcommand{\tw}{\text{w}}
\newcommand{\tv}{\text{v}}
\newcommand{\tN}{\text{N}}

\newcommand{\nn}{\nonumber}

\newcommand{\cref}{{\bf [check ref]}}

\newcommand{\cU}{{\cal U}}

\psfrag{n1}{$\nu_1$}
\psfrag{n2}{$\nu_2$}
\psfrag{n1'}{$\nu_1'$}
\psfrag{n2'}{$\nu_2'$}
\psfrag{n9}{$\nu_9$}
\psfrag{n10'}{$\nu_{10}'$}
\psfrag{t1}{$\tilde{\nu}_1$}
\psfrag{t2}{$\tilde{\nu}_2$}
\psfrag{t9}{$\tilde{\nu}_9$}
\psfrag{t1'}{$\tilde{\nu}_1'$}
\psfrag{t2'}{$\tilde{\nu}_2'$}
\psfrag{t10'}{$\tilde{\nu}_{10}'$}

\newcommand{\Na}{{\text{N}^a}}

\newcommand{\Nk}{{\mathcal{N}_\kappa}}
\newcommand{\Nl}{{\mathcal{N}_\lambda}}

\def\blfootnote{\xdef\@thefnmark{}\@footnotetext} 
\long\def\symbolfootnote[#1]#2{\begingroup%
\def\thefootnote{\fnsymbol{footnote}}\footnote[#1]{#2}\endgroup}

\begin{document}

\baselineskip=17pt

\begin{titlepage}
\begin{flushright}
\parbox[t]{1.8in}{
}
\end{flushright}

\begin{center}

\vspace*{ 1.2cm}

{\large \bf  The $\cN=1$ effective action of  F-theory compactifications 
\\[.1cm]   
}

\vskip 1.2cm

\begin{center}
 \bf{Thomas W. Grimm \footnote{grimm@th.physik.uni-bonn.de}}
\end{center}
\vskip .2cm

{\em Bethe Center for Theoretical Physics, Universit\"at Bonn, \\[.1cm]
Nussallee 12, 53115 Bonn, Germany}
 \vspace*{1cm}

\end{center}

\vskip 0.2cm
 
\begin{center} {\bf ABSTRACT } \end{center}

The four-dimensional $\cN=1$ effective action of F-theory compactified on 
a Calabi-Yau fourfold is studied by lifting a three-dimensional M-theory 
compactification. The lift is performed by using T-duality realized via a Legendre transform on the level of 
the effective action, and the application of vector-scalar duality 
in three dimensions. The leading order K\"ahler potential and 
gauge-kinetic coupling functions are determined.
In these compactifications two sources of gauge theories 
are present. Space-time filling non-Abelian seven-branes arise at the singularities of 
the elliptic fibration of the fourfold. Their couplings are included by 
resolving the singular fourfold. Generically a $U(1)^r$ gauge theory arises 
from the R-R bulk sector if the base of the elliptically fibered Calabi-Yau fourfold 
supports $2r$  harmonic three-forms. The gauge coupling functions depend 
holomorphically on the complex structure moduli of the fourfold, comprising 
closed and open string degrees of freedom. The four-dimensional electro-magnetic 
duality is studied in the three-dimensional effective theory obtained after 
M-theory compactification. A discussion of matter couplings transforming in the 
adjoint of the seven-brane gauge group is included.

\hfill August, 2010
\end{titlepage}

\tableofcontents
\newpage

%
%

\section{Introduction}

In connecting string theory with effectively four-dimensional observable 
physics one breaks the high amount of symmetry present in 
the ten-dimensional formulation on a compactification background. 
Demanding the existence of an effective four-dimensional $\cN=1$ supergravity theory 
improves the stability of the compactifications while still leading to interesting 
phenomenological scenarios.   
However, even if one is able to guarantee the presence 
of the observed particle spectrum, the symmetry breaking induces 
a massless moduli sector which would be in conflict with experiment. 
There has been vast progress to establish scenarios to stabilize 
these scalar fields \cite{Douglas:2006es,Blumenhagen:2006ci,Denef:2008wq}. This is particularly important 
due to the fact, that these fields determine the value of the four-dimensional 
couplings and scales. Global consistency conditions restrict valid scenarios, but 
are not believed to single out specific effective theories, or select a preferred vacuum.  
Thus, it is a crucial task to determine the realized $\cN=1$ 
effective theories with a realistic observable sector 
and evaluate their generic features.

In the last years there has 
been much progress in the study of Type II string 
compactifications with D-branes. These set-ups allow to localize 
non-Abelian gauge theories in the internal geometry, which arise on 
space-time filling D-branes or via light D-brane states on vanishing 
cycles in a singular geometry. Charged matter fields can 
arise on intersections of these branes. This 
allows for the possibility to construct set-ups  
resembling the four-dimensional particle 
spectrum and couplings of the MSSM \cite{Blumenhagen:2005mu,Blumenhagen:2006ci}. 
However, intersecting D-brane models do not naturally  
support GUT theories, since the required couplings are only generated 
at the non-perturbative level \cite{Blumenhagen:2009qh,Blumenhagen:2008zz}. Moreover, despite much progress, 
the implementation of moduli stabilization for D-brane deformations poses additional 
complications. Both of these issues are naturally addressed in F-theory 
compactifications which provides a geometrization of general seven-brane configurations.

In this work we will focus on Type IIB compactifications in which the 
gauge symmetries arises from space-time filling seven-branes or from
vector zero modes of the R-R four-form. 
Type IIB compactifications with general seven-brane sources
and varying dilaton-axion $\tau$ are known as F-theory compactifications \cite{Vafa:1996xn}. 
F-theory provides a geometrization of the seven-branes by 
considering backgrounds which admit two
extra dimensions confined on an auxiliary two-torus with complex
structure parameter $\tau$. The aim of this work is to study 
the four-dimensional effective supergravity actions arising 
in such F-theory compactifications. Demanding $\cN=1$ 
supersymmetry implies that F-theory has to be 
compactified on Calabi-Yau fourfolds $X_4$. The 
existence of the auxiliary two-torus then translates to 
the fact that $X_4$ has to be an elliptic fibration over some base $B_3$. 
The singularities of this fibration determine the 
four-cycles in $B_3$ wrapped by the seven-brane of Type IIB string 
theory on $B_3$. 

In general F-theory compactifications the elliptic 
fibration $X_4$ can admit singularities associated with exceptional 
groups. This fact permits the existence of 
four-dimensional models with induced couplings following 
the selection rules of the representation theory of exceptional groups. This is 
crucial for many minimal constructions of GUT models with unified gauge-groups 
$SU(5)$ or $SO(10)$.  
This fact has revived a recent interest in the construction of realistic GUT models in 
F-theory starting with \cite{Donagi:2008ca,Beasley:2008dc,Beasley:2008kw}.\footnote{In particular, phenomenological issues concerning the flavor structure  have been addressed 
in local F-theory models. An incomplete list of recent refs. is \cite{Heckman:2008qa,Hayashi:2009ge,Cecotti:2009zf,
Conlon:2009qq,Marchesano:2009rz,Hayashi:2009bt,Dudas:2009hu,King:2010mq}.} 
However, in the aim to connect these models with a moduli 
stabilizing sector the coupling to gravity is essential. Compact F-theory GUT 
models have been recently studied in \cite{Marsano:2009ym,Blumenhagen:2009yv,Grimm:2009yu,
Chen:2010ts,Chen:2010tp}, while F-theory up-lifts of orientifold models have been 
constructed in \cite{Collinucci:2008zs,Blumenhagen:2009up}.
In particular, 
the explicitly resolved Calabi-Yau fourfolds with a single non-Abelian gauge group of \cite{Blumenhagen:2009yv,Grimm:2009yu,Chen:2010ts} can be viewed 
as simple examples for which the effective action computed in this work can be evaluated.
A landscape of semi-realistic GUT models can thus be studied in an 
effective theory with a GUT sector coupled to a moduli stabilization sector.

Aiming to derive such a four-dimensional effective theory 
one immediately faces the problem that F-theory is not a fundamental theory 
with a twelve-dimensional weak coupling formulation. 
The detour which one has to take is to consider first a compactification 
of M-theory on the Calabi-Yau fourfold $X_4$ to three space-time 
dimensions \cite{Vafa:1996xn}. 
Using the fact that $X_4$ is an elliptic fibration, one 
then identifies a limit which shrinks the fiber torus and grows the 
fourth non-compact dimension \cite{Denef:2008wq}. The basic idea is 
to fiberwise apply the duality between M-theory on $T^2$ and Type IIB string theory on $S^1$
after applying T-duality \cite{Schwarz:1995dk,Aspinwall:1995fw}. Sending the volume of the $T^2$ to zero 
corresponds to the F-theory limit in which the $S^1$ becomes large. 
Since this will be our way to extract the four-dimensional 
effective action it will be crucial to formulate this limit very explicitly.
M-theory vacua on Calabi-Yau fourfolds have been analyzed in refs.~\cite{Becker:1996gj,Dasgupta:1999ss}.\footnote{In 
this paper we will neglect the effects of warping by working at very large compactification volume.} 
The effective three-dimensional action of M-theory on a general, non-singular Calabi-Yau fourfold 
at large volume has been studied in ref.~\cite{Haack:1999zv}.\footnote{See also refs.~\cite{Lust:2005bd,Braun:2008pz}, and references therein, for discussions on the F-theory action in compactifications with more supersymmetry.} We simplify the computations of \cite{Haack:1999zv} 
and provide the tools to implement the F-theory limit. 

To keep control of the M-theory reduction in the F-theory limit all Euclidean M5-branes wrapped on holomorphic 
six-cycles in $X_4$ have to be of large action. We argue that this requirement can be 
fulfilled even in the limit of a small elliptic fiber, and can be traced back to the fact that 
the appropriately identified vanishing $T^2$-volume $R$ 
is connected with the action of a corresponding Euclidean M5-brane on $B_3$ via a Legendre transform.
The F-theory limit can be obtained by following Euclidean M5-branes in M-theory 
which map to finite action Euclidean D3-branes in F-theory \cite{Witten:1996bn}.  
Furthermore, it will be crucial to use the existence of a well-defined four-dimensional 
theory arising after decompactification. 

The F-theory limit will be extended to 
set-ups with non-Abelian gauge symmetry on seven-brane in which case a more subtle scaling of the 
fields has to be applied. The four-dimensional gauge field degrees of freedom 
in the Coulomb branch arise in the three-dimensional action as vector modes of the 
M-theory three-form paired with blow-up modes of the singular fibration of $X_4$.
In three-dimensional theories with four supercharges vectors   
pair with real scalars into multiplets. 
We will use this fact and show that the four-dimensional K\"ahler potential and 
gauge-kinetic coupling function can be encoded by a single three-dimensional 
kinetic potential $\mathbf{K}^{\rm M}$. Expanding $\mathbf{K}^{\rm M}$
for small fiber volumes $R$ of the elliptic Calabi-Yau fourfold, one can  
directly read of the four-dimensional F-theory K\"ahler potential and 
gauge-coupling functions. In this way we show that for a non-Abelian 
gauge group the gauge-coupling function is to leading order given 
by the volume of the wrapped seven-branes and explore the 
structure of further corrections. We show that matter transforming 
in the adjoint of the seven-brane gauge group can be coupled in the 
M-theory reduction. In the F-theory lift this matter corresponds to 
deformations of the seven-branes as well as Wilson line degrees of freedom. 
The D-terms together with the flux induced contribution are computed 
using the four-dimensional K\"ahler metric along the gauged directions. 
Also the flux induced superpotential together with a direct coupling of the 
adjoint deformations and Wilson lines on the seven-branes are discussed. 

A second source of four-dimensional 
gauge symmetry arises from the scalars 
appearing as coefficients of an expansion of the M-theory 
three-form into a basis of non-trivial three-forms on $B_3$.
In three dimensions, one uses 
the complex structure of $X_4$ to combine these scalars 
into complex fields which span a complex torus bundle $\mathbb{T}$ 
over the moduli space of complex structure deformations of $X_4$. 
In three dimensions massless vectors and scalars are dual, and we 
show that these scalars indeed lift to four-dimensional vectors in the 
F-theory limit. The holomorphic gauge coupling functions dependents on the 
complex structure moduli of the fourfold $X_4$, and encodes the geometry 
of the torus bundle $\mathbb{T}$. 
Four-dimensional 
electro-magnetic duality can be studied in the three-dimensional 
theory and allows to constrain the form of the gauge-couplings 
and non-perturbative superpotentials.

The paper is organized as follows. In section~\ref{Fsystematics} we systematically 
introduce the lift of three-dimensional 
M-theory compactifications to four-dimensional F-theory compactifications. The 
K\"ahler moduli sector is discussed from a Type IIB and an M-theory perspective, and we introduce 
the usage of Legendre transforms in the effective theories. In section~\ref{non_Abelian}
we discuss non-Abelian seven-branes, and show how their couplings can be 
studied in an M-theory reduction. It will be crucial to resolve the singularities 
of the elliptic fibration and later define an appropriate scaling limit in the 
F-theory lift. The Abelian gauge theories arising from the R-R four-form of Type IIB 
string theory is introduced in section~\ref{RRsection} using the M- to F-theory lift. 
We discuss the action of electro-magnetic duality on the three-dimensional variables
and comment on the properties of the gauge coupling function and the three-dimensional 
superpotential. Finally, in section~\ref{matter_couplings}, matter transforming in the adjoint 
of the seven-brane gauge group is included. The corrections to the K\"ahler potential 
are determined in the F-theory lift. This allows to study the D-terms and to comment on the 
$\cN=1$ superpotential of the effective theory. 

For the convienience of the reader, the main equations and results of this work 
are summarized in appendix \ref{summary_app}.

\section{Systematics of F-theory compactifications} \label{Fsystematics}

In this section we summarize the general strategy which we use to study F-theory compactifications.
A first look at Calabi-Yau fourfolds 
with seven-branes in subsection \ref{TypeIIBbase} allows us to summarize the uncharged 
four-dimensional spectrum of a general F-theory compactification in subsection \ref{spectrum_summary}.
In subsection \ref{TypeIIBRemarks} simple aspects of the K\"ahler moduli space are discussed from 
a Type IIB perspective. 
Our main tool in the determination of the effective action will be to understand F-theory compactifications 
as a limit in the M-theory K\"ahler moduli space in which a new non-compact dimension grows. 
This lift from a three- to a four-dimensional compactification is introduced in 
subsection \ref{fromMtoF}, and will be extend in the following 
sections. Finally, in subsection \ref{complex_structure_section_1} we summarize some 
basics about complex structure deformations on smooth Calabi-Yau fourfolds.

\subsection{Type IIB string compactifications with seven-branes \label{TypeIIBbase}}

Recall that ten-dimensional Type IIB string theory is believed to admit a 
non-perturbative $Sl(2,\bbZ)$ symmetry. This group acts non-trivially 
on the dilaton-axion $\tau = C_0 + i e^{-\phi}$, where $C_0$
is the R-R axion and $e^{\langle \phi \rangle}=g_s$ is the string coupling. 
Since eight-dimensional branes couple to $\tau$ this 
implies that in addition to the well-known D7-branes also more general seven-branes 
obtained by an $Sl(2,\bbZ)$ transformation can be included in a consistent Type IIB 
compactification. 
In this work we will focus on four-dimensional Type IIB 
string theory compactified on a complex three-dimensional manifold $B_3$. 
The seven-branes are wrapped on four-cycles, i.e.~divisors, in $B_3$. 
In consistent solutions 
the tension of the seven-branes is locally canceled by the 
positive curvature of a K\"ahler base manifold $B_3$. 
All 7-branes are sources for $\tau$ 
and hence are identified by the behavior of the dilaton-axion $\tau$ profile on 
the compactification background. 
Close to the seven-branes $\tau$ can vary significantly and one is not 
generically at weak string coupling.   
The weak coupling limit is only approached when moving the 
seven-branes together to form O7-planes with the remaining branes being 
D7-branes~\cite{Sen:1996vd}.
In this limit the cancellation of seven-brane tension with the 
curvature of $B_3$ translates to the standard cancellation condition 
of the D7-brane tadpole.

Type IIB compactifications with general seven-brane sources
and varying complex dilaton-axion $\tau$ are
known as F-theory vacua \cite{Vafa:1996xn}. F-theory provides a 
geometrization of the seven-branes by considering backgrounds $\mathbb{M}_{3,1} \times X_4$ which admit
two auxiliary extra dimensions. Such extended solutions are 
constructed by attaching at each point of the original 
ten-dimensional Type IIB target space $\mathbb{M}_{3,1} \times B_3$ an auxiliary two-torus
with complex structure parameter $\tau$.
The profile of $\tau$ translates to the non-trivial torus fibration structure of the fourfold $X_4$
when moving along $B_3$. The supersymmetry conditions and equations of motion for 
$\tau$ enforce this fibration to be an elliptic fibration, with $\tau$ varying holomorphically in the 
complex coordinates $\underline{u}$ of $B_3$. Note that in order for the 
four-dimensional effective theory to be $\cN=1$ supersymmetric the fourfold $X_4$ has to be a
Calabi-Yau manifold. 

Examples of such fourfolds
can be represented by a complex polynomial constraints in a projective or toric 
ambient space. In particular, one can consider $X_4$ encoded by the Weierstrass equation 
\beq \label{Weierstrass}
   P_W = x^3 - y^2 + f(\underline{u}) x z^4 + g(\underline{u}) z^6 = 0\ ,
\eeq
as well as a number of additional constraints $P_{i}(\underline{u})=0$. 
The coordinates $(x,y,z,\underline{u})$ are in general 
restricted by a number of scaling relations. In particular, for $(y,x,z)$ one has the scaling relation $(y,x,z) \cong (\lambda^3 y,\lambda^2 x,\lambda z)$. Note that for $f$ and $g$ constant \eqref{Weierstrass} indeed defines a 
a two-torus given by a degree $6$ hypersurface in weighted projective space $\bbP_{3,2,1}$. This 
two-torus can degenerate over divisors in $B_3$. These degeneration loci 
precisely locate the seven-branes on the base, and are determined by 
the discriminate 
\beq \label{def-Delta}
   \Delta = 27\, g^2 + 4\, f^3\ .
\eeq
The dilaton-axion profile $\tau(\underline{u})$ is then specified 
by the value of the classical $SL(2,\bbZ)$ modular invariant $j$-function
$j(\tau) = 4(24\,f)^3/\Delta$.

In general $\Delta$ can factorize into several components corresponding 
to different intersecting seven-branes.
The singularities of the elliptic fibration over these seven-brane 
divisors in $B_3$ determine the gauge groups on the seven-branes. 
In the rest of the paper we will 
restrict to configurations with a single stack of seven-branes on a surface $\cS$ 
leading to non-Abelian gauge group $G$. More precisely, we will restrict to 
examples where the class of $\Delta$ splits as 
\beq \label{Delta_split}
   [\Delta ] = rk(G) [\cS] + [\Delta']\ ,
\eeq
where $rk(G)$ is the number of seven-branes wrapped on $\cS$.
Such a non-trivial factorization might be imposed by tuning the complex 
structure of a smooth $X_4$ to appropriately degenerate the elliptic 
fibration over $\cS$ to obtain the non-Abelian gauge symmetry.
A number of examples of this type have been constructed in 
refs.~\cite{Blumenhagen:2009yv,Grimm:2009yu}, with the aim to build compact $SU(5)$ GUT models.

It is important to note that in case one has a non-Abelian gauge group on $\cS$ 
the degeneration of the elliptic fibration 
is so severe that the Calabi-Yau fourfold itself
becomes singular. Let us denote this singular space by $X_4^{\rm sing}$.  
In this case it is not possible to work with the singular space $X_4^{\rm sing}$
directly since the topological quantities such as the Euler characteristic and 
intersection numbers are not well-defined. In many cases, however,  
the singularities can systematically be blown up to obtain a smooth 
geometry  $\hat X_4$ \cite{Candelas:1996su,Bershadsky:1996nh,Candelas:1997eh,Blumenhagen:2009yv,Grimm:2009yu}.
We will discuss this blow-up process in more detail in section \ref{Ell+seven}.
In summary, a possible way to construct examples is
\beq \label{res_deform}
  X_4 \quad \xrightarrow{\ \text{fix complex str.}\ } \quad X_4^{\rm sing}
     \quad \xrightarrow{\ \text{K\"ahler blow-up}\ } \quad \hat X_4\ .
\eeq
It is important to stress that many singular elliptic Calabi-Yau fourfolds $X_4^{\rm sing}$
might not admit a corresponding smooth $X_4$ obtained by complex 
structure deformation. In principle,
this does not mean that F-theory on such spaces is not defined. Examples 
which do not admit a $X_4$ have a minimal gauge-group and 
have been studied intensively (see e.g.~refs.~\cite{Candelas:1996su,Bershadsky:1996nh,Candelas:1997eh}, for Calabi-Yau 
threefold examples). 
For the discussion in this work, it will be necessary for $X_4^{\rm sing}$ 
to at least admit a resolution $\hat X_4$ as described in section \ref{Ell+seven}.

Using such a set-up, we can determine the spectrum of the four-dimensional effective 
theory. The precise number of zero modes will be determined by the topological data 
of the Calabi-Yau fourfold and the surface $\cS$ together with 
the non-trivial gauge-field configuration on $\cS$. To summarize this 
spectrum will be the task of the next subsection.

\subsection{The four-dimensional spectrum \label{spectrum_summary}}

In order to study F-theory compactifications, it is crucial to identify 
the fields which appear as the light degrees of freedom in the 
four-dimensional effective theory. In general, this is a hard task since 
the precise number does not only depend on the topological data of the 
Calabi-Yau fourfold $\hat X_4$, but also will require a knowledge of 
the flux background used in the reduction. To get a clue on the
spectrum we take the standard strategy to first consider the 
case where all background fluxes are switched off. However, 
this will particularly be problematic when determining the chiral 
spectrum from the seven-branes, since four-dimensional 
chirality is only induced by a non-trivial flux background on the 
branes. We will need to return to this issue in section \ref{matter_couplings}.
 
To summarize the spectrum in the absence of fluxes, we first summarize a 
few facts on the dimension of the cohomology groups of Calabi-Yau fourfolds 
$\hat X_4$. Let us denote by $h^{p,q}(B_3)$ and $h^{p,q}(\hat X_4)$
the Hodge numbers of the base $B_3$ and the resolved Calabi-Yau fourfold 
$\hat X_4$ respectively. Note that a Calabi-Yau fourfold has three independent 
non-trivial Hodge numbers $h^{1,1}(\hat X_4)$, $h^{2,1}(\hat X_4)$, 
and $h^{3,1}(\hat X_4)$. The remaining non-vanishing Hodge numbers are given
by 
\beq
  \hat X_4:\quad h^{4,0}=h^{0,0}=h^{4,4}=1\ ,\qquad h^{2,2} = 2(22 + 2 h^{1,1} 
  + 2 h^{3,1}- h^{2,1})\ .
\eeq
For the basis $B_3$ one only finds two non-trivial Hodge numbers $h^{1,1}(B_3)$
and $h^{2,1}(B_3)$. Note that the fact that $h^{i,0}(\hat X_4)=0,\, i=1,2,3$
implies that also $h^{i,0}(B_3)=0$.

In the absence of background flux the number of four-dimensional chiral multiplets 
with scalar components being a complex scalar can be given in terms of the Hodge numbers
of $X_4$, $\hat X_4$ and $B_3$. 
The number of chiral multiplets is given by
\beq
   n_{\rm c} = h^{3,1}(X_4) + h^{1,1}(B_3) + (h^{2,1}(\hat {X}_4) - h^{2,1}(B_3)) \ .
\eeq
To see this, one has to enter the precise prescription of the dimensional 
reduction. The first contribution $h^{3,1}(X_4)$ is readily seen to arise
from the deformations of the complex structure of $X_4$.\footnote{If a smooth $X_4$ 
does not exist, one can analyze $h^{3,1}(\hat X_4)$ complex structure deformations 
which preserve the singularities of $X_4^{\rm sing}$. Additional deformations can 
exist locally and are included as two-form variations on $\cS$ as discussed in section \ref{matter_couplings}.} These fields include 
the deformation moduli of the seven-branes and we will discuss this sector 
in subsection \ref{complex_structure_section_1}. The second contribution $h^{1,1}(B_3)$ arises   
from the zero modes of the K\"ahler form $J$ of $\hat X_4$ expanded in harmonic 
two-forms on $B_3$. These are complexified by scalars arising from 
the R-R four-form $C_4$ of Type IIB string theory. The third contribution 
$h^{2,1}(\hat {X}_4) - h^{2,1}(B_3)$ is harder to identify in the Type IIB context, since it involves the full 
fourfold $\hat X_4$. We will come back to the explanation in 
the later parts of this work. In the next sections we will stepwise consider 
more and more general compactifications with fourfolds $\hat X_4$ 
with non-trivial $h^{2,1}(\hat X_4),h^{2,1}(B_3)$ and identify the corresponding 
fields and their effective couplings.

Let us also display the equation for the number of vector multiplets in 
the four-dimensional spectrum. As we will discuss in more detail below, the number 
of possible $U(1)$ vector multiplets is given by  
\beq \label{number_nv}
  \tilde n_{\rm v} = (h^{1,1}(\hat {X}_4)-h^{1,1}(B_3)-1) + h^{2,1}(B_3) \ . 
\eeq
Here $n_{\rm v}$ includes the number of $U(1)$'s in the 
non-Abelian gauge group $G$ over $S$, as we recall in section \ref{non_Abelian}.
However, in the actual F-theory compactification the Coulomb branch of 
the seven-brane gauge theory is not accessible and these $U(1)$'s enhance 
to the full non-Abelian gauge connection of $G$.
Hence, the actual number of $U(1)$ vector multiplets in a four-dimensional 
F-theory compactification is  
\beq
  n_{U(1)} = \tilde n_{\rm v}- \text{rank}(G) \ ,
\eeq
with $\tilde n_{\rm v}$ as given in \eqref{number_nv}. 
Finally, it is straightforward to identify the $h^{2,1}(B_3)$ $U(1)$'s in \eqref{number_nv} 
which arise by expanding the R-R four-form into $h^{2,1}(B_3)$
harmonic three-forms.

\subsection{Remarks on the Type IIB dimensional reduction} \label{TypeIIBRemarks}

The aim of this work is to study the four-dimensional effective 
action of a Type IIB compactification with seven-branes as 
sketched in sections  \ref{TypeIIBbase} and \ref{spectrum_summary}.
More precisely, we aim to determine the $\cN=1$ characteristic data 
in the general four-dimensional supergravity action
\cite{Wess:1992cp}
\beq\label{N=1action}
  S^{(4)} = -\int \tfrac{1}{2}R_4 * 1 +
  K_{I \bar J} \cD M^I \wedge * \cD \bar M^{\bar J}  
  + \tfrac{1}{2}\text{Re}f_{\Lambda \Sigma}\ 
  F^{\Lambda} \wedge * F^{\Sigma}  
  + \tfrac{1}{2}\text{Im} f_{\Lambda \Sigma}\ 
  F^{\Lambda} \wedge F^{\Sigma} + V*1\ ,
\eeq
where the scalar potential is given by
\beq\label{N=1pot}
V=
e^K \big( K^{I\bar J} D_I W {D_{\bar J} \bar W}-3|W|^2 \big)
+\tfrac{1}{2}\, 
(\text{Re}\; f)^{-1\ \Lambda\Sigma} D_{\Lambda} D_{\Sigma}
\ .
\eeq
The complex fields $M^I$ are the bosonic fields of chiral multiplets, and 
might be gauged by vectors $A^\Lambda$ in the derivative $\cD M^I$. Such 
gaugings will lead to the appearance of D-terms $D_\Lambda$ in $V$ \eqref{N=1pot}. 
Note that $K_{I \bar J}$ and $K^{I \bar J}$ are the K\"ahler metric 
and its inverse, where locally one has $K_{I\bar J} = \partial_I \bar\partial_{\bar J} K(M,\bar M)$ with 
$\partial_I = \partial/\partial M^I$.
The scalar potential is expressed in terms of the K\"ahler-covariant 
derivative $D_I W= \partial_I W + 
(\partial_I K) W$.

In order to study the effective four-dimensional dynamics of an 
F-theory compactification one first might attempt to 
start with Type IIB supergravity and perform a
dimensional reduction on $B_3$. However, one immediately 
encounters the problem that the fields 
\beq \label{monodr-fields}
  \tau = C_0 + i e^{-\phi}\ ,\qquad G_2=C_2 - \tau B_2\ ,
\eeq
cannot be used in a Kaluza-Klein expansion, since they vary non-trivially 
over the threefold $B_3$. This is due to the fact that these 
fields transform under the $Sl(2,\bbZ)$ symmetry of Type IIB string theory 
as
\beq  \label{tau_trans}
  \tau \ \rightarrow \ \frac{a \tau + b}{c \tau  + d} \ ,\qquad G_2 \ \rightarrow \ \frac{G_2}{c\tau +d}\ ,
\eeq
where the matrix $(a,b|c,d)$ is an element of $Sl(2,\bbZ)$.
There will in general be effectively invariant modes on the 
fourfold $X_4$ and we will show in the following sections that one 
can systematically perform a Kaluza-Klein reduction of F-theory via M-theory. 

Type IIB string theory, however, also admits D3-branes which couples to the R-R four-form $C_4$.
Neglecting the induced lower-dimensional brane charges on their world-volume, these D3-branes 
are invariant under the $Sl(2,\bbZ)$ transformations. Using such D3-brane probes one thus is able 
to study at least part of the four-dimensional effective action. For example,
let us consider an Euclidean D3-brane wrapped on some divisor $D^{\rm b}_\alpha$
of $B_3$. At large volume of $B_3$ this brane will have a classical instanton 
action 
\beq \label{Talpha_simple_base}
   T_\alpha^{\rm b} = \frac{1}{2 \ell_s^4} \int_{D^{\rm b}_{\alpha}} J_{\rm b} \wedge J_{\rm b} + i \int_{D^{\rm b}_{\alpha}} C_4\ ,
\eeq
where $J_{\rm b}$ is the K\"ahler form on $B_3$ in the ten-dimensional Einstein frame in units $\ell_s^2$. To render $T_\alpha^{\rm b}$
dimensionless one has to multiply the first term by $\ell_s^{-4}$. 
Clearly, this expression is only valid up to corrections 
in the fields \eqref{monodr-fields}. A key observation is, that the volumes of four-cycles
together with the R-R four-form $C_4$ appear as complex fields in the 
four-dimensional $\cN=1$ effective theory \cite{Giddings:2001yu,Grimm:2004uq}. The volumes $v_{\rm b}^\alpha$ of two-cycles in the 
usual expansion $J_{\rm b} = v^\alpha_{\rm b} \omega_\alpha$, with a basis $\omega_\alpha$ of $H^2(B_3,\bbZ)$, 
are in fact scalars in dual linear multiplets $(v^\alpha,C^{(2)}_\alpha)$, where the $C^{(2)}_\alpha$  
are two-forms (see ref.~\cite{Grimm:2004uq} for a detailed discussion). 
This general fact implies that $T_\alpha^{\rm b}$ can be obtained by a 
Legendre transformation from a \textit{kinetic potential} $\tilde K^{\rm F}$ as 
\beq \label{general_Talpha_base}
  T_\alpha^{\rm b} =  \partial_{L^\alpha_{\rm b}} \tilde K^{\rm F} + i \int_{D_\alpha^{\rm b}} C_4\ , \qquad L^\alpha_{\rm b} = \frac{v^\alpha_{\rm b}}{\cV_{\rm b}} \ ,
\eeq
where $\cV_{\rm b}$ is the quantum volume of $B_3$.
In general, the kinetic potential $\tilde K^{\rm F}$ 
depends on $L^\alpha$ and the other complex fields $M^I$ of the effective theory. It is related 
to the K\"ahler potential $K^{\rm F}$ via
\beq \label{def-KF}
  K^{\rm F} (T,\bar T|M,\bar M) = \tilde K^{\rm F}  - \tfrac12 (T^{\rm b}_\alpha + \bar T^{\rm b}_\alpha) L^\alpha_{\rm b} \ , 
\eeq
where the right-hand side has to be evaluated as a function of $T_\alpha^{\rm b}$ by solving \eqref{general_Talpha_base}
for $L^{\alpha}_{\rm b}$. Clearly, to reproduce the simple form \eqref{Talpha_simple_base} with $L^\alpha_{\rm b}$ 
as given in \eqref{general_Talpha_base}, one has to take 
\beq \label{simple_KFtilde}
  \tilde K^{\rm F} = - 2 \log \cV_{\rm b}= \log \big(\tfrac{1}{3!} L^\alpha_{\rm b} L^\beta_{\rm b} L^\gamma_{\rm b} \cK_{\alpha \beta \gamma} \big)\ ,\qquad \cV_{\rm b} = \frac{1}{3!} \int_{B_3} J_{\rm b}\wedge J_{\rm b} \wedge J_{\rm b}
\eeq
where $\cK_{\alpha \beta \gamma}$ is the triple intersection of three divisors $D_\alpha, D_{\beta}, D_\gamma$.
Note that the simple forms  \eqref{Talpha_simple_base}, \eqref{simple_KFtilde} of $T^{\rm b}_\alpha$, $K^{\rm F}$
are obviously not complete. This can already be inferred by comparing these expressions 
with the ones obtained in the orientifold picture \cite{Giddings:2001yu,Grimm:2004uq}. 
There various known corrections to 
both $T^{\rm b}_{\alpha}$ and $K^{\rm F}$, which do, however, depend on the dilaton-axion $\tau$.
It will be crucial to understand how these corrections are captured in an F-theory
compactification.
One of the aims of this paper is to compute part of these corrections to 
the K\"ahler potential $K^{\rm F}$ (or $\tilde K^{\rm F}$), and $T_\alpha^{\rm b}$ including 
the other light fields arising as complex structure deformations, seven-brane moduli, matter fields, and fields 
from the combinations of the form \eqref{monodr-fields}. In order to
deal with varying dilaton-axion $\tau$ it thus will be necessary to 
analyze the higher-dimensional geometry $\mathbb{M}_{3,1} \times X_4$ which turn out to 
be tractable in an M-theory framework.

\subsection{From M-theory to F-theory \label{fromMtoF}}

In this subsection we describe four-dimensional F-theory vacua 
as special limits of M-theory compactifications. To begin with, let us recall 
the basic steps to link M-theory and Type IIB string theory \cite{Denef:2008wq}. 
Consider M-theory compactified on a two-torus $T^2$, naming one of the one-cycles, the A-cycle, and 
the other one-cycle, the B-cycle. The metric background is thus of the form 
\beq \label{metric_11d}
   ds^2_{11}=\frac{v^0}{\I\, \tau} \big((d x + \R\, \tau d y)^2 + (\I\, \tau)^2 d y^2 \big) + ds^2_9 \ ,
\eeq
where $\tau$ is the complex structure modulus of the $T^2$, and $v^0$ describes its volume. 
If the volume $v^0$ of the two-torus is small, one can pick one of the one-cycles, 
say the A-cycle, to obtain type IIA string theory. 
T-duality along the B-cycle leads to the corresponding Type IIB set-up, and identifies 
$\tau = C_0 + i e^{-\phi}$. One can then decompactify the T-dualized B-cycle to grow an 
extra non-compact direction.

This construction can be applied fiber-wise for the elliptically fibered Calabi-Yau 
fourfold $X_4$. Hence, one considers M-theory on $X_4$, which leads to a \textit{three-dimensional} 
theory with $\cN=2$ supersymmetry. The reduction and T-duality on the elliptic fiber  
leads to Type IIB string theory on $B_3 \times S^1$, where $B_3$ is the base. 
A fourth non-compact dimension is grown by decompactifying the $S^1$. Due to the T-duality operation
this limit corresponds to sending the $T^2$ volume $v^0 \rightarrow 0$.
One thus finds the duality 
\bea \label{F-theorylift}
  \text{M-theory on}\ X_4 \quad &\rightarrow& \quad \text{Type IIB on}\ B_3 \times S^1\ \text{with varying}\ \tau  \nn  \\
  \text{M-theory on}\ X_4 \ \text{with}\ v^0\rightarrow 0  \quad &\rightarrow& \quad  \text{F-theory on}\ X_4 
\eea
Let us note that this duality can only be performed in such a simple way, at points of the 
fibration were the two-torus does not degenerate. If singularities appear, one has 
to carefully identify the corresponding M-theory and F-theory non-perturbative constituents. 
In particular, singularities of the elliptic fibration in M-theory lead to Kaluza-Klein monopoles 
descending to Type IIA six-branes, while they yield seven-branes in the Type IIB set-up. 

From the M-theory perspective it is not surprising that the theory can be trusted 
for a varying $\tau$, since this parameter simply encodes the complex structure of an 
actual two-torus in the eleven-dimensional space. The expansion parameters kept small 
in the dimensional reduction will turn out to be the inverse volumes 
of the six-cycles in the Calabi-Yau fourfold $X_4$. 
Despite the fact that a complete formulation of M-theory 
is not known, one can attempt to include corrections which are known either 
via duality or in specific limits. For example, this indirect approach is used 
in the study of compactifications on singular fourfolds $X_4^{\rm sing}$, where 
M2 branes on vanishing cycles are believed to complete enhancements 
to non-Abelian gauge groups. However, also the limit from M-theory to F-theory is
in general subtle. Note that a supersymmetric M-theory compactifications
demands that one works with a Ricci flat metric on the Calabi-Yau fourfold. These 
metric properties are inherited by Type IIB on $B_3$, and it is non-trivial that an 
extension to the fiber directions exists. Fortunately, we will not need to make use 
of an explicit metric on $X_4$ or $B_3$. Nevertheless even on the level of cohomology 
and deformations a non-trivial mixing of open and closed string degrees of freedom will arise.

To make some first steps in analyzing the effective action, let us consider M-theory on a non-singular 
Calabi-Yau fourfold $X_4$ with an elliptic fibration. Since $X_4$ is smooth there will be no non-Abelian gauge 
symmetries in three dimensions. On such a fibration there is a natural set of divisors which 
span $H_6(X_4,\bbR)$. Firstly, one has the section of the fibration which is homologous to the 
base $B_3$. Secondly, there is the set of \textit{vertical divisors} $D_\alpha$ which are 
obtained as $D_\alpha = \pi^{-1}(D_\alpha^{\rm b})$, where $D_\alpha^{\rm b}$ is 
a divisor of $B_3$ and  $\pi$ is the projection to the base $\pi: X_4 \rightarrow B_3$.
For these smooth elliptic fibrations one has $h^{1,1}(B_3)=h^{1,1}(X_4)-1$ such divisors.
One can now attempt to use a probe M5-brane to analyze K\"ahler coordinates. 
At large volume the naive action for an M5-brane on such a $D_\cA =(D_0,D_\alpha)$ 
then reduces as 
\beq \label{Talpha_simple_Mtheory}
 T_\cA = \frac{1}{6}\int_{D_\cA} J \wedge J \wedge J + i \int_{D_\cA} C_6 \ ,
\eeq
where $v^0$ is the volume of the elliptic fiber. 
In order to make $T_\cA$ dimensionless 
one would need to multiply the first term in \eqref{Talpha_simple_Mtheory} with $\ell_M^{-6}$.
We will suppress units in most of the equations below.
To determine the K\"ahler potential $K^{\rm M}$ 
for the fields $T_0,T_\alpha$  one analyses the Weyl rescaling to the three-dimensional 
Einstein frame. In a large volume compactification, only the classical volume 
$\cV$ arises as pre-factor of the Einstein-Hilbert term. Comparing this with
the $e^{K^{\rm M}}$ pre-factor in the scalar potential, on infers \cite{Haack:1999zv} 
\beq \label{class_KM}
   K^{\rm M} = - 3 \log \cV\ , \quad \qquad \cV = \frac{1}{4!}\int_{X_4} J \wedge J \wedge J \wedge J\ .
\eeq
To evaluate $K^{\rm M}$ as a function of $T+\bar T$ one first expands the K\"ahler form $J$
as
\beq \label{J_exp_1}
  J=v^0 \omega_0 + v^\alpha \omega_\alpha\ .
\eeq
where $\omega_0,\omega_\alpha$ are the two-forms Poincar\'e dual to $B_3,D_\alpha$.
Using this expansion on has to solve \eqref{Talpha_simple_Mtheory} for 
the modes $v^0,v^\alpha$ of $J$ and insert the result into \eqref{class_KM}.
This evaluation is more conveniently performed in a 
dual picture, which we explain in more generality next.

In general, one notes that the  
$v^\alpha,v^0$ appear as elements of vector multiplets $(v^\alpha,A^\alpha)$ and $(v^0,A^0)$
with the vectors arising in the expansion of the M-theory three-form $C_3$ as
\beq \label{C3-exp_simple}
  C_3 = A^0 \wedge \omega_0 + A^\alpha \wedge \omega_\alpha\ .
\eeq 
Hence, again one expects the $T_\cA =(T_0,T_\alpha)$
to be given by 
\beq \label{TM_general}
    T_\cA  = \partial_{L^\cA} \tilde K^{\rm M} + i\rho_\cA\ ,
\eeq
where $\rho_\cA$ generalize the imaginary parts in \eqref{Talpha_simple_Mtheory}, and $L^\cA = (R,L^\alpha)$ are defined as
\beq \label{def-LA}
   R = \frac{v^0}{\cV}\ ,\qquad  L^\alpha = \frac{v^\alpha}{\cV} \ .  
\eeq
This is the analog of \eqref{general_Talpha_base}, but now for some M-theory kinetic potential $\tilde K^{\rm M}$, and 
$\cV$ being the quantum volume of $X_4$. Again, the K\"ahler potential $K^{\rm M}$ is related to 
the kinetic potential via the Legendre transform
\beq \label{Legendre1}
     K^{\rm M} (T,\bar T|M,\bar M) = \tilde K^{\rm M}  - \tfrac12 (T_\cA + \bar T_\cA) L^\cA \ , 
\eeq
where $M^I$ are other complex scalars in the three-dimensional theory. It is straightforward 
to evaluate $\tilde K^{\rm M}$ at large volume to obtain 
the simple expression \eqref{Talpha_simple_Mtheory} for $T_\cA=(T_0,T_\alpha)$. One finds
\beq \label{simple_KMtilde}
  \tilde K^{\rm M} = \log (R) +  \log \big(\tfrac{1}{3!} L^\alpha L^\beta L^\gamma \cK_{\alpha \beta \gamma} + \tfrac{1}{2} R L^\alpha L^\beta \cK_{\alpha \beta}  + \tfrac{1}{2} R^2 L^\alpha \cK_{\alpha} + R^3 \cK \big)\ .
\eeq
The intersection numbers we introduced are 
\beq \label{def-Kabc}
    \cK_{\alpha \beta \gamma} = B_3 \cdot D_\alpha \cdot D_\beta \cdot D_\gamma = \int_{B_3} \omega_\alpha \wedge \omega_\beta \wedge \omega_\gamma\ ,
\eeq
and similarly for $\cK_{\alpha \beta} = \cK_{0 0\alpha \beta}$ and the remaining terms. Note that
for an elliptic fibration the intersection numbers satisfy 
\beq
  \cK_{\alpha \beta \gamma \delta} = D_\alpha \cdot D_\beta \cdot D_\gamma \cdot D_\delta = 0\ ,
\eeq
for the vertical divisors. This allows us to split of the factor $\log (R)$ in \eqref{simple_KMtilde}.
Furthermore, it implies that for an elliptic fibration one has
\beq
  T_0 = \frac{1}{R}+ p(R) + i \rho_0\ ,\qquad  T_\alpha = \frac{v^0}{2} \int_{D_\alpha^{\rm b}} J \wedge J+ i \rho_\alpha\ ,
\eeq
where $p(R)$ is a power-series in $R$ regular in the limit $R\rightarrow 0$.
Observe that $\R T_0$ starts precisely with an inverse of $R$, with $R$ being proportional to the 
volume of the elliptic fiber $v^0$ as introduced in \eqref{metric_11d} and \eqref{J_exp_1}. This will be 
the key to study the F-theory limit.

Let us now turn to the discussion of the F-theory limit. This limit will identify the 
M-theory compactification on $X_4$ to three dimensions with a four-dimensional F-theory 
compactification. Let us consider Type IIB string theory on $S^1\times B_3$.  
Before taking the limit there is one distinguished dimension which 
corresponds to one of the torus directions in the elliptic fiber of $X_4$. 
Labeling this fourth dimension by $x^3$ the Type IIB metric is of the 
form 
\beq \label{4d-metric_red}
   ds^2_{\rm IIB} = r^{-2} g_{\mu \nu}^{3} dx^\mu dx^\nu + r^2 (dx^3 + A_{\mu}^0 dx^\mu)^2 + ds^2_{B_3}\ ,  \qquad \mu,\nu=0,1,2\ ,
\eeq
where $r$ is the radius of the fourth dimension, $g_{\mu \nu}^3$ is the 
three-dimensional Einstein frame metric, and $A_\mu^0$ is a three-dimensional vector.
As recalled above, the F-theory limit is obtained by performing the reduction and T-duality, 
and sending $r\rightarrow \infty$ to decompactify the fourth dimension. Note
that $A_0$ in \eqref{4d-metric_red} is identified in the M-theory to F-theory 
lift with the vector $A_0$ in the multiplet $(R,A_0)$ introduced in \eqref{C3-exp_simple} and \eqref{def-LA}.
Also $A_0$ in the dimensional reduction with metric \eqref{4d-metric_red} is in a vector multiplet 
$(r^{-2},A_0)$. Hence, one identified the radius $r$ in \eqref{4d-metric_red} with $R$ in \eqref{def-LA}
as 
\beq \label{Rr_id}
    R = r^{-2}\ .
\eeq
One thus realizes that the shrinking of the elliptic fiber $R\rightarrow 0$ corresponds 
to growing an extra dimension. Furthermore, due to the Legendre transform from $R$ to $T_0$
one sees that $R\rightarrow 0$ corresponds to $\R T_0 \rightarrow \infty$. This pushes the
analysis into a regime, where Euclidean M5-branes wrapped on the base $B_3$ become very massive
and do not correct the $\cN=1$ data \cite{Witten:1996bn}. 

The F-theory lift can be studied further by realizing that M5-branes on vertical divisors 
will turn into D3-branes wrapped on the four-cycles $D_\alpha^{\rm b}$ in $B_3$ 
with finite action \cite{Witten:1996bn}. Hence, in the F-theory limit, one indeed identifies 
the $T_\alpha^{\rm b}$ introduced in section \ref{TypeIIBRemarks} and $T_\alpha$ of section \ref{fromMtoF}, given e.g.~in \eqref{Talpha_simple_base} and \eqref{Talpha_simple_Mtheory}.
Equivalently, one can use the fact that $L^\cA$ in the M-theory reduction are elements of
vector multiplets. Since, $T^{\rm b}_{\alpha}$ introduced in \eqref{TM_general} remains 
finite in the F-theory limit, also $L^{\alpha}$ defined in \eqref{def-LA} has to remain 
finite. Hence, we conclude that the F-theory limit is more accurately given by 
\beq \label{F-theory-limit}
   R = \ell_M^{6} \cdot \frac{v^0}{\cV}\quad  \rightarrow \quad 0 \ ,  \qquad \quad L^\alpha = \ell_M^6 \cdot \frac{v^\alpha}{\cV}\quad  \rightarrow \quad L^\alpha_{\rm b} = \ell_s^4 \cdot \frac{v^\alpha_{\rm b}}{\cV_{\rm b}}\quad \text{finite}\ , 
\eeq
or, equivalently, by 
\beq
    \R T_0 \quad  \rightarrow \quad \infty\ , \quad \qquad \qquad \qquad T_\alpha  \quad  \rightarrow \quad T_{\alpha}^{\rm b} \quad \text{finite} \ ,\qquad \quad\ 
\eeq
where we have restored the $\ell_s$ and $\ell_M$ dependence to elucidate the limit. 
Let us stress that \eqref{F-theory-limit} implies a non-trivial scaling of the 
$v^\alpha$. Taking the volume $\cV$ to be quartic in $v^0,v^\alpha$ one finds
for $v^0 \propto \epsilon$ that $v^\alpha \propto 1/\sqrt{\epsilon}$ and $\cV \propto 1/\sqrt{\epsilon}$ in the limit $\epsilon \rightarrow 0$. Using \eqref{F-theory-limit} for a quartic $\cV$ the elliptic fiber volume $R$ 
scales in the F-theory limit as $R \cong \cV_{\rm b}^2/\cV^3$. 
The claim is that the limit and identification \eqref{F-theory-limit} also holds if one 
includes further corrections and other moduli. In this case, however, one has to replace $\cV$ with the 
appropriate quantum volume, as we will discuss below.
 
The key objects we will study in the following sections 
are the two three-dimensional kinetic potentials $\tilde K^{\rm M}$ 
and $\mathbf{K}^{\rm M }$ and their four-dimensional lifts.
The latter potential $\mathbf{K}^{\rm M }$ is obtained from $\tilde K^{\rm M}$ via 
a Legendre transform of \textit{only} the vector multiplets $(L^\alpha,A^\alpha)$, since these 
vectors lift to four-dimensional chiral multiplets $T_\alpha^{\rm b}$. 
$\mathbf{K}^{\rm M}$ is given by 
\beq \label{kin_F-dual1_preview}
  \mathbf{K}^{\rm M} = \tilde K^{\rm M} - \tfrac12(T_\alpha + \bar T_\alpha) L^\alpha\ ,
\eeq
where $L^\alpha$ is replaced by its Legendre transform $\R T_\alpha = \partial_{L^\alpha} \tilde K^{\rm M}$.
This will be discussed in more detail in section \ref{seven_brane_gaugelift}.
Even in the presence of vector multiplets and further complex scalars, we will argue that the full 
M-theory kinetic potential $\mathbf{K}^{\rm M}$ admits in the limit \eqref{F-theory-limit} the expansion 
\beq \label{KM_claim}
   \mathbf{K}^{\rm M} = \log R + K^{\rm F} - \frac{1}{R} g + \cO(R)\ ,
\eeq
where $K^{\rm F}$ is the K\"ahler potential of the four-dimensional F-theory compactification 
and the real function $g$ will encode the dynamics of four-dimensional vector fields. 
The expression \eqref{KM_claim} is readily checked when inserting the kinetic potential 
$\tilde K^{\rm M}$ given in \eqref{simple_KMtilde} into \eqref{kin_F-dual1_preview}, together 
with $K^{\rm F}$ given in \eqref{def-KF}, \eqref{simple_KFtilde}. Clearly, in this simple 
case one has $g=0$.

Let us remark that in principle one should explicitly evaluate the M-theory K\"ahler potential around
the F-theory limit \eqref{F-theory-limit}, including corrections to the large volume expressions.
To some extend this is indeed possible by using mirror symmetry for Calabi-Yau fourfolds. The basic strategy is to 
construct the mirror $X'_4$  to $X_4$ and rewrite the K\"ahler potential $K^{\rm M} = - 3\log \cV$ 
using the mirror periods of the mirror $(4,0)$ form $\Omega'$ using the techniques described in subsection \ref{complex_structure_section_1}.
In other words, one needs to compute 
\beq \label{KX'}
K^{\rm M} = - 3 \log \int_{X'_4} \Omega' \wedge \bar \Omega'\ .
\eeq
However, it is important to stress, that this potential has to be restricted to a \textit{real submanifold} of dimension
$h^{1,1}(X_4)$ in the mirror complex structure moduli space. This is analog to the description of Type IIA orientifolds \cite{Grimm:2004ua}.
Moreover, $K^{\rm M}$ will need to be evaluated as a function of the coordinates $T_{\cA}$, with real 
parts given by the real parts of certain $\Omega'$ periods. Giving a precise formulation of this 
mirror identification is beyond the scope of this work. However, let us note that certain corrections 
have already been computed in \cite{Grimm:2009ef}. In particular, it was shown that the Calabi-Yau 
fourfold volume $\cV$ of $X_4$, appearing in \eqref{class_KM}, is corrected by the terms  
\beq \label{Delta_cV}
   \Delta \cV = \frac{5 \zeta(4)}{2^4 (2\pi i)^6}\int_{X_4} c_2(X_4)^2 +  k_1 \int_{X_4} J \wedge c_3(X_4) + k_2 \int_{X_4} J^2 \wedge c_2(X_4) + \ldots\ ,
\eeq 
where $c_2,c_3$ is the second and third Chern class of $TX_4$, and $k_i$ are some numerical constants. 
In particular, there is no correction proportional 
to the Euler number $\chi(X_4)$.\footnote{This is in contrast to the case of Calabi-Yau threefolds, where the 
Euler number of the threefold corrects the threefold volume.} This seems to be crucial when taking the F-theory limit
in the corrected $\cN=1$ coordinates $T_\cA$.

\subsection{Complex structure deformations} \label{complex_structure_section_1}

In this section we recall some basic facts about the complex structure deformations of a
non-singular Calabi-Yau fourfold $\hat X_4$ mainly following \cite{Greene:1993vm,Mayr:1996sh,Klemm:1996ts}. 
Around a fixed background complex structure these
arise as metric deformations with purely holomorphic and anti-holomorphic indices
\beq \label{deform_4}
  \delta g_{\bi \bj} = -\frac{1}{3 ||\Omega ||^2} \bar \Omega_{\bi}^{ \ klm}  
  (\chi_{\cK})_{klm\bj}  \, \delta z^\cK \ ,
\eeq
where $\Omega$ is holomorphic $(4,0)$-form on $\hat X_4$. 
Hence, the complex structure deformations 
are counted by the basis $\chi_\cK, \cK=1,\ldots, h^{3,1}(\hat X_4)$ 
of $H^{3,1}(\hat X_4)$. As for Calabi-Yau threefolds the infinitesimal 
deformations $\delta z^\cK$ are unobstructed in the absence of background 
fluxes and can be extended to a complex $h^{3,1}(\hat X_4)$-dimensional moduli space $\cM^{\rm cs}$. 
The metric on this moduli space is given by 
\beq \label{def-Kcs_gen}
  G_{\cK \bar \cL} = \partial_{z^\cK}\partial_{\bar z^\cL} K^{\rm cs}= \frac{\int_{\hat X_4} \chi_\cK \wedge \bar \chi_\cL }{\int_{\hat X_4} \Omega \wedge \bar \Omega} \ ,\qquad \qquad K^{\rm cs} = - \log \, \int_{\hat X_4} \Omega \wedge \bar \Omega \ ,
\eeq
where we also recalled that $G_{\cK \bar \cL} $ is K\"ahler and is thus locally given by 
the derivative of the K\"ahler potential $K^{\rm cs}$.

The K\"ahler potential $K^{\rm cs}$ for the complex structure deformations $z^\cK$ can be expressed through the 
periods of $\Omega$ as we discuss momentarily. It is important to stress that 
in the fourfold case the variations of the $(4,0)$ form $\Omega$ with respect to the complex structure 
deformations do 
not span the full cohomology $H^4(\hat X_4)$, but 
rather only a subspace $H_H^4(\hat X_4)$, known as the primary horizontal subspace of $H^4(\hat X_4)$ \cite{Greene:1993vm}.
It takes the form
\beq \label{horiz}
  H^4_H(\hat X_4,\bbC) = H^{4,0} \oplus H^{3,1} \oplus H^{2,2}_H  \oplus H^{1,3} \oplus H^{0,4}\ ,
\eeq
where $H^{2,2}_H$ consists of the elements in $H^{2,2}$ which 
can be obtained as second variation of $\Omega$ with respect to the 
complex structure on $X_4$.\footnote{The fact that not all $H^{4}(\hat X_4)$ can be reached as variation 
of $\Omega$ is in contrast to the Calabi-Yau threefold case. In 
the threefold case one can simply define the periods of the holomorphic $(3,0)$-form
by introducing an integral homology basis of $H_3(Y_3,\bbZ)$.}  
In the fourfold case, however, one has 
to introduce a special basis $\gamma^{(i)}_{a}$ of $H_4^H(\hat X_4,\bbC)$ which 
inherits the integrality properties of a mirror dual basis of $\oplus_q H^{q,q}(\hat X'_4,\bbZ)$ \cite{Greene:1993vm},
where $\hat X'_4$ is the mirror of $\hat X_4$.
This allows to define the periods  
\bea \label{def-periods}
    \Pi^{(i)\, a_i}  & = &  \int_{\gamma^{(i)}_{a_i}} \Omega \ , \qquad \quad i = 0,\ldots,4\ ,\qquad a_i=1,\ldots,h^{4-i,i}_H(\hat X_4)\ , \\
    {\Pi} &\equiv& (\Pi^{(0)},\Pi^{(1)a},\Pi^{(2)\alpha },\Pi^{(3)a},\Pi^{(4)})\equiv (X^0,X^a,\cG^\alpha,\cF^a,\cF^0)\ ,
\eea
where $h^{4-i,i}_H(\hat X_4)$ denote the dimensions of the respective cohomologies in \eqref{horiz}.
The basis $\gamma^{(i)}_{a_i}$ can be chosen to admit the intersections 
\beq  \label{int_on4}
  \Sigma \equiv \big({\gamma}^{(i)}_{a_i} \cap {\gamma}^{(j)}_{b_j} \big) = \left( 
  \begin{array}{ccccc}
   0 & 0 & 0 & 0 & 1	\\
   0 & 0 &0& \eta & 0 \\
   0 & 0 & Q & 0 & 0\\
   0 & \eta^{T} &0 &0 &0\\
   1 & 0 & 0 & 0 & 0    
\end{array} \right)
\eeq
which only yields for $j=4-i$ the non-zero intersection matrices $\eta_{a b},Q_{\alpha \beta}$. The 
group preserving $\Sigma$ will be denoted by $G_\Sigma$:
\beq \label{def-GSigma}
   N \in G_\Sigma: \quad N^T \Sigma N = \Sigma\ .
\eeq
Inserting 
\eqref{def-periods} and \eqref{int_on4} into \eqref{def-Kcs_gen} the K\"ahler potential $K^{\rm cs}$ can 
be expressed in terms of the periods $\Pi$ using
\beq
  \int_{\hat X_4} \Omega \wedge \bar \Omega = \Pi^T \Sigma \bar{\Pi} = X^{0} \bar \cF^{0} + \eta_{ab} X^a \bar \cF^b + Q_{\alpha \beta} \cG^\alpha \bar \cG^\beta + c.c\  .
\eeq
Using the $(p,q)$-structure of forms obtained as derivative of $\Omega$ one 
derives a number of vanishing conditions which 
translate into non-trivial conditions on the periods $\Pi$.\footnote{See, e.g., refs.~\cite{Mayr:1996sh,Denef:2004ze} 
for a summary of these conditions.}
However, in contrast to the Calabi-Yau threefold case there does not exist a  
prepotential which determines the periods.

In principle the periods $\Pi$ can be computed explicitly 
as a function of the complex structure deformations $z^\cK$ by 
methods discussed, for example, in refs.~\cite{Greene:1993vm,Mayr:1996sh,Klemm:1996ts}. 
More precisely, 
given specific constraints \eqref{Weierstrass} which determine 
$X_4$ as a hypersurface or complete intersection in a 
toric or projective ambient space, one can compute 
a set of differential equations, the  Picard-Fuchs equations, 
which admit a linear combination of the $\Pi$ as solution.
The precise linear combination of the solutions to Picard-Fuchs equations 
at a given point in moduli space $\cM^{\rm cs}$ can be fixed by analytic continuation 
and the analysis of monodromies around special loci 
in $\cM^{\rm cs}$ (see \cite{Candelas:1990rm} for the original work on Calabi-Yau threefolds, and \cite{Grimm:2009ef}
for extensions to Calabi-Yau fourfolds). While technically rather involved this gives, 
at least for Calabi-Yau fourfolds with few complex structure moduli, a 
prescription to compute $K^{\rm cs}$ explicitly at various points in the 
moduli space for a given $\hat X_4$.

In the computation of the periods $\Pi$ it is crucial 
to have a detailed understanding of the \textit{global structure} 
of the moduli space $\cM^{\rm cs}$. As already mentioned this structure is 
largely captured by the monodromies around the special loci, such as the 
fourfold conifold \cite{Grimm:2009ef}\footnote{The deformed fourfold conifold is also known as 
Stenzel space \cite{Stenzel}.}, the 
large complex structure point, and the orbifold locus. More precisely one 
has to determine the monodromy matrices $M$ and their generated group $G^{\rm sym}$
\beq \label{monodromy_group}
  \Pi \ \rightarrow \ M \Pi , \quad  M \in G^{\rm sym} \subset G_{\Sigma}\ ,
\eeq
when encircling the special loci of $\cM^{\rm cs}$. $G^{\rm sym}$ is typically 
only defined via the specifications of its generators $M$, and encodes the global 
symmetries of $\cM^{\rm cs}$.

Let us end this section by noting that the complex structure deformations of $\hat X_4$ 
can be obstructed when allowing for non-trivial background fluxes \cite{Douglas:2006es,Blumenhagen:2006ci,Denef:2008wq}. 
More precisely, in an M-theory reduction on $\hat X_4$, the complex structure moduli of $\hat X_4$ are 
obstructed by a non-trivial flux background $G_4$ appearing in 
the Gukov-Vafa-Witten superpotential \cite{Gukov:1999ya}
\beq \label{W_GVW}
  W = \int_{\hat X_4} \Omega \wedge G_4\ .
\eeq
This superpotential can be computed explicitly by evaluating the periods $\Pi$ 
in an integral basis \cite{Grimm:2009ef,Jockers:2009ti}.
However, this is only the correct $W$ for a compactification on a non-singular $\hat X_4$ 
of M-theory. In the F-theory limit the superpotential \eqref{W_GVW} will be further refined due to 
the appearance of non-Abelian gauge symmetries at singularities of $X^{\rm sing}_4$.

\section{Non-Abelian seven-branes in F-theory compactifications} \label{non_Abelian}

In this section we systematically include non-Abelian gauge 
groups into the discussion of the four-dimensional F-theory effective action. 
Recall that in compactifications with multiple seven-branes on a 
divisor $\cS$ in $B_3$ the gauge-theory on their world-volume 
will be a non-Abelian group $G$. 
In the following we will concentrate on simply laced gauge groups which reside in the 
ADE series and can be obtained by singularities of the elliptic fibration
of a Calabi-Yau fourfold $X_4^{\rm sing}$. Hence, we will concentrate on groups 
$SU(N),SO(N)$ and the exceptional groups $E_6,E_7,E_8$. More precisely, we 
consider a stack of seven-branes on the divisor $\cS$ on the 
base. We denote by $F = dA + A \wedge A$ the eight-dimensional field-strength on 
their world-volume $\cW=\mathbb{M}_{3,1} \times \cS$. The gauge field 
transforms in the adjoint of the group $G$, and 
an overall $U(1)$-factor might split off, as familiar for the case $U(N)=SU(N)\times U(1)$. Such 
$U(1)$ factors often play a special role, and can be included in the 
analysis of the effective action as described in \cite{Grimm:2010ez}.
In order to determine the effective four-dimensional theory one splits $F$ into
contributions with two four-dimensional indices, mixed indices and two indices on $\cS$:
\beq \label{F-exp}
   F =F_4    + F_w + F_{\rm flux}\ .
\eeq 
The modes of the first set $F_4$ correspond to four-dimensional gauge fields 
and will be discussed in more detail in subsections \ref{non-Abel_M} and \ref{seven_brane_gaugelift}.
The second set $F_w$ in \eqref{F-exp} are defined 
to be one-forms on $\cS$ and one-forms in $\mathbb{M}_{3,1}$ and hence capture 
the Wilson line degrees of freedom as discussed in subsection \ref{Wilson}. 
The last set $F_{\rm flux}$ captures non-trivial flux configurations on the seven-branes 
as briefly discussed in section \ref{matter_couplings}. Note that in an F-theory compactification 
the underlying group theory is actually encoded geometrically, due to the presence 
of the singularities of the fibration over $\cS$. In subsection \ref{Ell+seven}
we discuss that this remains to be the case after the resolution of these singularities.

\subsection{Singularity resolutions for seven-brane gauge theories} \label{Ell+seven}

Recall that the elliptic fibration will be singular 
over the discriminant $\Delta$ given in \eqref{def-Delta}.
If the discriminant $\Delta$ factorizes the components will 
correspond to different seven-branes. As already noted in \eqref{Delta_split}
we will restrict to the case that $\Delta$ has two components 
$[\Delta] = rk(G) [\cS] + [\Delta']$. Here $rk(G)$ is the rank 
non-Abelian gauge group on the $rk(G)$ seven-branes wrapped on $\cS$.
The gauge groups over the divisors $\cS$ can be determined explicitly using 
generalizations of the Tate formalism \cite{Bershadsky:1996nh}. 
Let us split the basis of vertical divisors $D_\alpha$, $\alpha = 1,\ldots h^{1,1}(B_3)$ introduced in section~\ref{Fsystematics} as
\beq \label{Dalpha_split}
   D_\alpha = (S,D_\beta')\ .
\eeq
This split will be convenient in the following, since $S$ plays a distinguished 
role in the analysis of the gauge symmetries.

In case of non-Abelian gauge groups the 
elliptic fibration and the Calabi-Yau fourfold $X_4^{\rm sing}$ itself
becomes singular as in \eqref{res_deform}.  
The singularities can systematically be blown up to obtain a smooth 
geometry \cite{Candelas:1996su,Bershadsky:1996nh,Candelas:1997eh,Blumenhagen:2009yv,Grimm:2009yu}. In the 
following we will consider cases where $X_4^{\rm sing}$ admits a split simultaneous resolution
\beq \label{blow-up_space}
   \pi:\quad  \hat {X}_4 \ \rightarrow \ X_4^{\rm sing}\ ,
\eeq
where $\pi$ is the blow-down map from the smooth fourfold $\hat {X}_4$ to $X_4^{\rm sing}$.
The existence of a split simultaneous resolution implies that one can identify 
rank$(G)$ irreducible divisors $\hat D_i$ describing the resolution of the 
ADE singularity over $\cS$. Each of these divisors is a $\bbP^1$ bundle over 
$\cS$, and the $\hat D_i$ intersect at generic points in $\cS$ 
as the Dynkin diagram of $G$ in the elliptic fiber of $\hat {X}_4$.
Note that it is important 
to shift 
\beq \label{def-S'}
   S \ \rightarrow \ S' = S + \sum_i a^i \hat D_j\ , 
\eeq
where $a^i$ are the Dynkin numbers associated to the Dynkin node $\hat D_i$.
In the following we will use the basis
\beq \label{def-Dalpha_new}
    D_\alpha = (S',D_\beta')\ ,
\eeq
which, by a slight abuse of notation, replace the $D_\alpha$ introduced in 
\eqref{Dalpha_split}.
The redefinitions \eqref{def-S'} and \eqref{def-Dalpha_new} are 
performed to ensure that the intersection numbers satisfy
\beq \label{Ki_vanish}
  \cK_{i \alpha \beta \gamma} \equiv \hat D_j \cdot D_\alpha \cdot D_\beta \cdot D_\gamma = 0 \ .
\eeq
We will show below, that the existence of such a condition in an 
appropriately chosen basis is in accord with constraints of four-dimensional 
$\cN=1$ supersymmetry.
On the resolved geometry of the ADE singularity one then has 
\beq \label{hatDhatD}
   (\hat D_{i} \cdot \hat D_{j} + C_{ij} \, S' \cdot B_3) \cdot D_\alpha \cdot  D_\beta  = 0 \ ,
\eeq
for all vertical divisors $D_\alpha$ in \eqref{def-Dalpha_new}. Here 
$C_{ij}$ is the Cartan matrix of $G$. 
Furthermore, one can now include in the intersection form \eqref{hatDhatD} the extended node
\beq \label{def-extendednode}
    \hat D_0 = S' - \sum_i a^i \hat D_i = S \ ,
\eeq
such that 
\beq \label{hatDhatD_ext}
  (\hat D_{i} \cdot \hat D_{j} + C_{ij} \, S' \cdot B_3) \cdot D_\alpha \cdot  D_\beta  = 0 \ , \qquad i,j=0,...,rk(G)\ ,
\eeq
where now $C_{ij}$ is the Cartan matrix of the extended Dynkin diagram. 
If $\hat {X}_4$ is realized as
a hypersurface or complete intersection in a projective/toric ambient 
space, a resolution of the ambient space itself can 
consistently resolve the enhanced ADE singularities 
on $\cS$ \cite{Candelas:1996su,Candelas:1997eh,Blumenhagen:2009yv}.
In these examples one can check explicitly by computing the intersection
numbers that \eqref{hatDhatD} and \eqref{hatDhatD_ext} are satisfied in the properly  
chosen topological phase.

The equations \eqref{hatDhatD}, \eqref{Ki_vanish} and \eqref{hatDhatD_ext} can 
also be written in terms of two-forms which are Poincar\'e dual
to $B_3,\hat D_i,D_\alpha$. We list them for completeness:\\[-1cm]
\begin{itemize}
  \item[1.] A two-form $\omega_0$ Poincar\'e dual to the base $B_3$ of the elliptic 
        fibration. \\[-1cm]
 \item[2.] Two-forms $\omega_\alpha, \alpha = 1,\ldots h^{1,1}(B)$ on $X_4$
        which are Poincar\'e dual to vertical divisors $D_\alpha = \pi^{-1}(D_\alpha^{\rm b})$, as
        in section \ref{fromMtoF}.  \\[-1cm]       
  \item[3.] Two-forms $\tw_i, i = 1,\ldots,\text{rank}(G)$ which are Poincar\'e dual to the blow-up divisors 
        $\hat {D}_i$ which were introduced after \eqref{blow-up_space}. The two-form $\tw_0$ 
        corresponding to the extended node of the Dynkin diagram $\hat D_0$ can be canonically 
        included in the discussion \cite{Grimm:2010ez,GKPW}. $\tw_0$ is not linearly independent of the two-forms 
        so far, and thus is not needed to form a basis of $H^{1,1}(\hat X_4)$. 
\end{itemize}
The intersection conditions \eqref{hatDhatD} and \eqref{hatDhatD_ext} translate to  
\beq \label{tw_identity}
  \cK_{ij \alpha \beta} \equiv \int_{\hat X_4} \tw_i \wedge \tw_j \wedge \omega_\alpha \wedge \omega_\beta
   = - C_{ij} \int_{\cS} \omega_\alpha \wedge \omega_\beta \equiv -C_{ij}\, \cK_{S|\alpha \beta}\ ,
\eeq
where we have used in the second equality that $\omega_0,\omega_{S'}$ are Poincar\'e dual to $B_3,S'$. 
The last equality defines the intersection form $\cK_{S|\alpha \beta}$ on $\cS$ of the 
$D_\alpha$ of the ambient space. It is a well-known fact that there can be additional non-trivial 
two-cycles on $\cS$ which are not induced from intersections of $\cS$ with the $D_\alpha$.
These elements will be included later on, but do not alter the intersection analysis presented here.
 
It is important to stress that the singularity of the elliptic fibration can 
vary over $\cS$ and enhance to groups larger than $G$ along
complex curves and points. In this case the resolution of the singularity becomes 
more involved, as discussed in detail in refs.~\cite{Andreas:1999ng}. 
However, if $\hat {X}_4$ is realized as
a hypersurface or complete intersection in a projective/toric ambient 
space, a resolution of the ambient space itself can 
consistently resolve the enhanced ADE singularities at all co-dimensions on $\cS$ \cite{Candelas:1996su,Candelas:1997eh,Blumenhagen:2009yv,Grimm:2009yu}. In this case one 
systematically resolves the ambient space with divisors $\tilde D_i, i=1,\ldots,rk(G)$ which at generic points 
of $\cS$ restrict to the resolving divisors $\hat D_i$ introduced in this subsection.  
In accord with the analysis of the effective action, this resolution increases the number of 
two-forms on $\hat X_4$ by $rk(G)$ forms $\tw_i$. The resolution of further enhancements along 
curves and points does not change $h^{1,1}(\hat X_4)$. However, one will find new four-forms on 
$\hat X_4$ which are not a wedge of two $(1,1)$-forms. These new four-forms are crucial in 
defining the $G_4$ fluxes determining the chiral spectrum.

\subsection{Non-Abelian gauge groups in M-theory \label{non-Abel_M}}

To study the F-theory lift we first discuss  the appearance of 
non-Abelian gauge groups  
in the M-theory picture. In order to do that, we recall that in M-theory 
$U(1)$ vector fields can arise from the three-form $C_3$ with field strength
$G_4 = dC_3$. Considering M-theory on the 
resolved fourfold $\hat X_4$ will correspond to 
the Coulomb branch of the gauge theory where 
$G$ is broken to $U(1)^{rk(G)}$ over $\cS$. To study the Coulomb 
branch one replaces the general expression \eqref{F-exp} 
by an expansion into the Cartan generators $\cT_i$ of the adjoint representation 
of $G$. We choose these to obey
\beq
   \text{Tr}(\cT_i \cT_j) = C_{ij}\ , \qquad \quad i,j=1,\ldots,rk(G)\ ,
\eeq
The expansion \eqref{F-exp} translates into components of the 
four-form $G_4$ by writing schematically  
\beq \label{G4lift}
  F=F^i\, \cT_i\qquad \rightarrow \qquad G_4|_{\rm brane} \cong F^i \wedge \tw_i = (F^i_4   + F^i_{\rm flux} + F_w^i) \wedge \tw_i  \ .
\eeq
Note that this cannot be a precise statement, since the M-theory reduction is to three rather then 
four space-time dimensions. It will be the task of the next subsection to make the lift more explicit.
We start with the vectors with field strength $F^i_4$, for which the lift \eqref{G4lift} 
is most directly applicable.

To obtain the complete set of massless three-dimensional 
vectors let us consider the expansion three-form $C_3$ into two-forms of $H^{1,1}(\hat X_4)$ as 
\beq \label{C3expansion_1}
   C_3|_{\rm vector} = A^0 \wedge \omega_0 + A^\alpha \wedge \omega_\alpha + A^i \wedge \tw_i\ ,
\eeq
where $A^0,A^\alpha,A^i$ are three-dimensional vector fields. 
In the three-dimensional $\cN=2$ theory these vectors are 
combined with the real scalars arising in the expansion of the 
K\"ahler form $J$ of $\hat X_4$ into vector multiplets. One expands
\beq \label{J-expansion}
   J = v^0 \omega_0 + v^\alpha \omega_\alpha + \tv^i \tw_i\ ,
\eeq
where the $\tv^i$ measure the volumes of the blow-up $\bbP^1$'s. 
As noted above, the $A^i$ are the $r$ $U(1)$ vector fields which correspond to
the Cartan generators of the non-Abelian 
gauge group $G$. The non-Abelian gauge symmetry arises in the limit 
in which the volumes of the blow-ups go to zero. Then M2 branes 
wrapped on chains of resolving $\bbP^1$ fibers become massless and 
provide the missing gauge degrees of freedom to form a non-Abelian gauge group $G$.
We will return to the restoration of $G$ in the discussion of the F-theory lift in 
the next section.

Let us next discuss the effective action of three-dimensional vector multiplets 
$(A^0,v^0)$, $(A^i,\tv^i)$ and $(A^\alpha,v^\alpha)$. Since we will be 
more general later on, we will include in the following expressions also 
a number of three-dimensional chiral multiplets $M^I$. 
As in \eqref{def-LA} we first introduce the rescaled variables which actually appear in 
the three-dimensional $\cN=2$ vector multiplets,
\beq \label{def-RxiL}
   R = \frac{v^0}{\cV}\ , \qquad \quad \xi^i = R \cdot \zeta^i =  \frac{\tv^i}{\cV} \ ,\qquad \quad L^\alpha = \frac{v^\alpha}{\cV}\ ,
\eeq
where we have defined also $\zeta^i$ by splitting off a factor of $R$.\footnote{This split will be convenient 
in the discussion of the F-theory lift.} Note that in the K\"ahler cone one generically has $\xi^i \leq 0 $ 
as $\tv^i$ appears in front of a two-form Poincar\'e dual to an exceptional 
divisor $\hat D_i$ in \eqref{J-expansion}. This ensures, in particular, that the volumes of the 
divisors $\hat D_i$ are positive as we will see below. 
The three-dimensional action 
for the vector multiplets $(A^{\hat \Lambda},\xi^{\hat \Lambda}) \cong (A^0,R),(A^i,\xi^i),(A^\alpha,L^\alpha)$
is of the form 
\bea\label{kinetic_lin_gen_1}
  S^{(3)} &=& \int-\tfrac{1}{2}R_3 *\mathbf{1} - 
  \tilde K_{I \bar {J}}\, dM^I \wedge * d \bar M^{J}
  + \tfrac{1}{4} \tilde K_{\hat \Lambda \hat\Sigma}\, 
  d\xi^{\hat \Lambda}\wedge * d\xi^{\hat \Sigma} \nn\\ 
  && - \tfrac{1}{4} \tilde K_{\hat \Lambda \hat\Sigma}\, F^{\hat\Lambda} \wedge * F^{\hat\Sigma}
     + \,  F^{\hat \Lambda} \wedge \I (\tilde K_{\hat\Lambda I}\, dM^I)\ ,
\eea
where the kinetic terms of the vectors and scalars are determined by 
a single real function, the kinetic potential $\tilde K(M^I,\bar M^J|\xi^{\hat \Lambda})$
with 
\beq
  \tilde K_{I {\bar J}} = \partial_{M^I} \partial_{\bar M^J} \tilde K \ , \qquad
  \tilde K_{\hat \Lambda \hat \Sigma} = \partial_{\xi^{\hat \Lambda}} \partial_{\xi^{\hat \Sigma}} \tilde K\ ,\qquad 
  \tilde K_{\hat \Lambda I} = \partial_{\xi^{\hat \Lambda}} \partial_{M^I} \tilde K \ .
\eeq
In the M-theory reduction we will denote the kinetic potential by $\tilde K^{\rm M}$
as above.

In a next step we aim to find the leading kinetic potential which 
captures the new degrees of freedom extending $\tilde K^{\rm M}$ 
given in \eqref{simple_KMtilde}.  
 One still has $\tilde K^{\rm M}=-3 \log \cV + K^{\rm cs}$, but 
now one needs to include the fields $\xi^i$ in the volume expansion. 
They appear as 
\beq \label{tildeKM1}
   \tilde K^{\rm M} = \log[\tfrac{1}{6} R L^\alpha L^\beta L^\gamma \cK_{\alpha \beta \gamma} - \tfrac{1}{4} \xi^i \xi^j \, C_{ij} \, L^\alpha L^\beta \cK_{S|\alpha \beta}+\cO(R^3,\xi^3)] + K^{\rm cs}(z)\ ,
\eeq
with $K^{\rm cs}$ given in \eqref{def-Kcs_gen}.
In this expression we have used \eqref{def-Kabc} for $\cK_{\alpha \beta \gamma}$, and 
\eqref{tw_identity} to obtain the term involving $\cK_{S|\alpha \beta}$. The expansion in 
the logarithm \eqref{tildeKM1} does not contain a linear term in $\xi^i$ as ensured by 
\eqref{Ki_vanish} on the resolved $\hat X_4$. This turns out to be crucial for 
the M-theory compactification to lift to a four-dimensional theory with gauge group on $\cS$.
Let us stress that the expression \eqref{tildeKM1} is a large volume expression, since it simply arose by expanding 
the quadruple intersections on a Calabi-Yau fourfold. Various corrections to this expression are expected as 
discussed briefly in section \ref{fromMtoF}. However, by simply using \eqref{tildeKM1}
and performing a Taylor expansion for small $R,\xi^i$ one nevertheless finds 
non-trivial match with the three- and four-dimensional expectations from the 
gauge theory and the gravity background. It would be 
very interesting to understand this match in more detail and 
to extend the following considerations to include further corrections.

To prepare for the discussion of the F-theory limit, let us 
now Taylor expand the above results for small $R,|\zeta^i| = |\xi^i|/R$.
The three-dimensional gauge kinetic coupling function in \eqref{kinetic_lin_gen_1} 
can be determined as the second derivative of $\tilde K^{\rm M}$ with respect to 
the scalars which are in three-dimensional vector multiplets. In particular, 
one finds that 
\beq \label{KijM}
  \tilde K_{ij}^{\rm M} = - \frac{C_{ij}\, \cK_{S| \beta \gamma} L^\beta L^\gamma}{2\, R \cV_L}  + \cO(R,\zeta^2)\ ,\qquad
   \cV_L \equiv \tfrac{1}{6}\cK_{\alpha \beta \gamma}L^\alpha L^\beta L^\gamma\ .   
\eeq
Furthermore, using $\tilde K^{\rm M}$ one readily evaluates the real parts of the 
dual K\"ahler coordinates $T_\alpha,T_i,T_0$ via \eqref{TM_general}.
Firstly, we have
\beq \label{special_Talpha}
  \R T_\alpha  =  \frac{1}{2} \frac{\cK_{\alpha \beta \gamma} L^\beta L^\gamma}{\cV_L} + \cO(R)\ , 
\eeq 
where we recall from \eqref{def-Dalpha_new} that one of these coordinates is $T_{S'}$ corresponding to the divisor $S'$.
Furthermore, one evaluates by using \eqref{tildeKM1} and \eqref{def-RxiL} that
\beq \label{specil_Ti}
  \R T_i = -  \frac{C_{ij}\zeta^j\,  \cK_{S |\beta \gamma} L^\beta L^\gamma}{2\,\cV_L} + \cO(R)\ , \qquad \quad 
  \R T_0 = \frac{1}{R} +\cO(R,\zeta^2) \ .
\eeq
Note that in the K\"ahler cone one has $\zeta^i \leq 0$ and one has $\R T_i \geq 0$, and $\R T_\alpha \geq 0$, $\R T_0 \geq 0$.

The expressions for $T_\alpha,T_i$  
exactly match the expectations from the point of view of a reduction from a four-dimensional 
to a three-dimensional gauge theory as discussed in \cite{Affleck:1982as,Seiberg:1996nz,Katz:1996th}.
To make this more precise, we will denote by $T_{S}$ the complex scalar
corresponding to the divisors $S \subset X_4$  introduced in \eqref{def-S'}, \eqref{def-extendednode} such that  
\beq
    T_{S} =  T_{S'} - \sum_{i=1}^{rk(G)} a^i T_i  \ .
\eeq
One can qualitatively analyze the three-dimensional superpotential 
obtained in the M-theory compactification from M5-brane instantons wrapped 
on divisors in $\hat X_4$. In ref.~\cite{Katz:1996th} it was argued that M5-branes 
on the blow-up divisors $\hat D_i$ as well as the extended node $\hat D_0$, defined 
in \eqref{def-extendednode}, satisfy the necessary criteria \cite{Witten:1996bn} to yield a non-trivial instanton 
correction to the superpotential. In addition also an M5-brane wrapped on the base $B_3$ satisfies 
these criteria \cite{Witten:1996bn}. Hence, the superpotential is expected to contain the terms
\beq \label{WM_exp}
   W^{\rm M} = \sum_{i=1}^{rk(G)} \cA_i e^{-T_i} +  \cB e^{-T_{S'}+ a^i T_i} + \cC e^{-T_0}\ ,
\eeq
where $\cA_i,\cB,\cC$ generically depend holomorphically on the other complex scalars, e.g.~the complex structure 
deformations, of the compactification. The terms with pre-factors $\cA_i$ correspond to the gauge theory instantons discussed 
in ref.~\cite{Affleck:1982as,Seiberg:1996nz}, while the term with $\cB$ is associated with a four-dimensional gauge instanton \cite{Seiberg:1996nz}
and vanishes in the 3d limit $r^2 = 1/R \rightarrow 0$. 
To see this one identifies $\tilde K^{\rm M}_{ij} \propto - 1/g_3^2$, where $g_3$ is the 
three-dimensional gauge coupling constant. 
Comparing \eqref{KijM} 
with \eqref{special_Talpha},\eqref{specil_Ti} one then finds $\R T_i \propto - \xi^i /g_3^2 $, and 
$\R T_{S'} \propto R / g_3^2 = 1/(r^2 g_3^2)$.\footnote{The expression for $\R T_S$ appears
to differ by a factor of $r$ from the results of \cite{Seiberg:1996nz,Katz:1996th}. However, 
this difference is readily explained by noting that in our analysis we work with a 
three-dimensional action Weyl rescaled to the three-dimensional Einstein frame with 
canonically normalized Einstein-Hilbert term. In refs.~\cite{Seiberg:1996nz,Katz:1996th} 
such a Weyl rescaling was not performed, which explains the different dependence on $r$. 
Also note that we have set $M_p = 1$ in three dimensions.} 
Finally, one can also interpret the last term in \eqref{WM_exp}. One 
first notes that $\R T_0 \propto r^{2}$. Following the duality described in 
section~\ref{fromMtoF} one notes that this instanton correction is due to a universal 
gravitational instanton in a four-dimensional theory on $\mathbb{M}_{2,1} \times S^1$ \cite{DGK}. 
In fact, the M5-brane becomes an NS5-brane in going from M-theory to Type IIA. This NS5-brane 
T-dualizes into a four-dimensional Taub-NUT geometry. The gravitational action 
is indeed proportional to $r^2$ \cite{Gibbons:1979nf,Emparan:1999pm}. 

\subsection{Seven-brane gauge theory in the F-theory lift} \label{seven_brane_gaugelift}

It is crucial to stress that in F-theory limit one necessarily takes 
the limit in which the resolution of the singularities is blown down. 
In other words, while one is able to access the Coulomb branch in M-theory 
this is no longer possible in F-theory. One thus has to extend 
the F-theory limit \eqref{F-theory-limit} to include the blow-up volumes. 
This leads us to replace \eqref{F-theory-limit} by 
\bea \label{F-theory-lift2}
  &\text{singular:} &\qquad R\ \rightarrow \ 0 \ ,\qquad \quad \ \ \zeta^i = \xi^i /R \ \rightarrow\ 0 \ , \\
  &\text{finite:} &\qquad L^\alpha \ \rightarrow \ L^\alpha_{\rm b}\ , \nn 
\eea
or
\beq
  \R T_0 \ \rightarrow \ \infty\ , \qquad \quad \R T_{i} \ \rightarrow \ 0 \ , \qquad \quad T_\alpha \ \rightarrow \ T_\alpha^{\rm b}\ .
\eeq
This implies the scaling of the $v^\cA$ as $v^0 \propto \epsilon$, $\tv^i \propto \epsilon^2$ and $v^\alpha \propto 1/\sqrt{\epsilon}$
in the limit $\epsilon \rightarrow 0$. 
In other words, the low-energy expansion of the F-theory 
effective action is around the special point \eqref{F-theory-lift2} in the K\"ahler moduli 
space. However, the variations $ \delta R, \delta \zeta^i$ are not un-physical but rather will appear as
degrees of freedom in four-dimensional fields. In particular, four-dimensional vector $A^i_{4}$ fields  
are given by 
\beq
  A^i_{4} = (A^i + \delta \zeta^i A^0,\delta \zeta^i)\ , 
\eeq
containing $\zeta^i$ defined in \eqref{def-RxiL}.
The lift of the three sets of three-dimensional vector multiplets is: 
 $(v^0,A^0)$ lifts to the metric of the fourth non-compact dimension $g_{33},g_{3\mu}$, 
 $(v^\alpha,A^\alpha)$ lift to chiral multiplets in four dimensions, and $(v^i,A^i)$ 
lift to $U(1)$ vectors corresponding to the Cartan generators of $G$. 
This is summarized in table \ref{F-theory_lift1}.

\begin{table}[h!] 
\begin{center}
\begin{tabular}{|c||c|c|} \hline  
 \rule[-0.3cm]{0cm}{0.9cm}
       3-dim~multiplet  & \multicolumn{2}{c|}{4-dim~F-theory}    \\ \hline
 \rule[-0.3cm]{0cm}{0.9cm}
 \multirow{2}{2cm}{$(L^{\cA},A^\cA)$}  & \quad $h^{1,1}(X_4)-h^{1,1}(B_3)-1$ vector mult. \quad & \quad $(\xi^i,A^i)\rightarrow A_4$\ \text{adjoint} \hspace*{.2cm} \\ \cline{2-3}
 \rule[-0.2cm]{0cm}{0.8cm}
 &  $h^{1,1}(B_3)$ chiral multiplets & $(L^\alpha,A^\alpha)\rightarrow T_\alpha$ \\ \cline{2-3}
 \rule[-0.2cm]{0cm}{0.8cm}
 &  extra dimension & $(R,A^0)\rightarrow (g_{33},g_{\mu 3})$ \\ 
 \hline  
\end{tabular} 
\caption{\textit{The F-theory lift of the fields arising 
from the K\"ahler form of $\hat X_4$.}} \label{F-theory_lift1}
\end{center}
\end{table}

To proceed we first recall that the multiplets $(v^\alpha,A^\alpha)$ actually 
lift to four-dimensional chiral multiplets. It is therefore convenient 
to rather work with $\mathbf{K}^{\rm M}(T_\alpha|\xi,R)$ given by 
\beq \label{kin_F-dual1}
  \mathbf{K}^{\rm M}(T_\alpha +\bar T_\alpha, M^I|\xi^i,R) = \tilde K^{\rm M} - \tfrac12(T_\alpha + \bar T_\alpha) L^\alpha\ ,
\eeq
where $L^\alpha$
is replaced by its Legendre transform
\beq \label{Talpha_Leg1}
   T_\alpha = \partial_{L^\alpha} \tilde K^{\rm M} + i \int_{D_\alpha} C_6 \ .
\eeq
Note that we have included additional scalars $M^I$ in the expressions for $\mathbf{K}^{\rm M}$ and 
$T_\alpha$, since their inclusion does not change the discussion presented here. 
These complex scalars are specified later, and include complex structure moduli, 
Wilson line moduli, matter fields, etc.
The expressions \eqref{Talpha_Leg1} and \eqref{kin_F-dual1} are very similar to the 
discussion to \eqref{TM_general} and \eqref{Legendre1}. However, it is crucial to 
stress, that in $\mathbf{K}^{\rm M}$, we have kept the vector multiplets containing $R,\xi^i$, 
and only dualized the multiplets $(v^\alpha, A^\alpha)$.
It is straightforward to check that 
\beq \label{KM_id}
   \frac{\partial \mathbf{ K}^{\rm M}}{\partial T_\alpha }= - \frac12 L^\alpha \   \ , \qquad 
\frac{\partial \mathbf{K}^{\rm M}}{\partial M} = \frac{\partial \tilde K^{\rm M}}{\partial M}\ , \qquad M \in (M^I,\xi^i,R)\ .
\eeq
Note that the right-hand sides of these expressions are evaluated by  first taking derivatives of $\tilde K^{\rm M}$ viewed 
as a function of $(L^\alpha,\xi^i,R)$ and $M^I$, and then use \eqref{Talpha_Leg1} to express the 
result as a function of $T_\alpha,R,\xi^i,M^I$. 
Note that by differentiating \eqref{Talpha_Leg1} one also finds
\beq \label{usefull1}
  \frac{\partial L^\alpha}{ \partial T_\beta} =  \tilde K^{{\rm M}\ L^\alpha L^\beta} \ ,\qquad   
  \frac{\partial L^\alpha}{\partial M} = - \tilde K^{{\rm M}\ L^\alpha L^\beta} \partial_M \tilde K^{\rm M}_{L^\beta} \ , \qquad M \in (M^I,\xi^i,R)\ .
\eeq

In order to study the F-theory lift one has to evaluate the kinetic potential $\mathbf{K}^{\rm M}$
in the limit \eqref{F-theory-lift2}. One thus performs a Taylor expansion of 
$\mathbf{K}^{\rm M}$ for small $R,\xi^i$ around the strict F-theory limit $R=\xi^i=0$.
In the following we will denote the restriction of a function $f(R,\xi)$ to the 
F-theory limit by $f|_*$.
The reasoning that there exists a four-dimensional $\cN=1$ supergravity theory in the F-theory 
limit  \eqref{F-theory-lift2} significantly restricts the form of the 
expansion of $\mathbf{K}^{\rm M}$. Namely, from a reduction of a four-dimensional
$\cN=1$ supergravity theory to three dimensions, as recalled in appendix \ref{scalar_vec_duality}, one 
infers that the kinetic potential has to admit the form
\beq \label{KM_claim2}
   \mathbf{K}^{\rm M} = \log R + K^{\rm F}(T,M) -\frac{1}{2 R}\, \R f_{ij}(T,M)\, \xi^i \xi^j + \ldots\ ,
\eeq
which resembles \eqref{d=3_red_potential} in appendix \ref{scalar_vec_duality}.
Here $K^{\rm F}(T,M)$ is the four-dimensional K\"ahler potential depending 
on the complex scalars $T_\alpha,M^I$, and $f_{ij}(T,M)$ is the four-dimensional 
holomorphic gauge coupling function. Note that the two terms involving 
$R$ are precisely the singular terms in the limit $R \rightarrow 0$. Further 
terms in the $(R,\xi^i)$-expansion will not be relevant in our computation of the four-dimensional 
effective theory. 
The expression \eqref{KM_claim2}  implies that 
\beq \label{firstvanish}
  \partial_{\xi^j}\mathbf{K}^{\rm M}\big|_* = \partial_{\xi^j}\tilde K^{\rm M}\big|_* = 0 \ ,
\eeq
in a set-up which consistently lifts to a four-dimensional $\cN=1$ F-theory compactification.
Similarly, comparing \eqref{KM_claim2} with the general Taylor expansion, one concludes that the 
four-dimensional gauge-coupling function is given by  
\bea \label{gauge_coupling1}
  \R\, f_{ij} &=& - R \cdot \mathbf{K}^{\rm M}_{\xi^i \xi^j}\big|_* = - R \cdot \big( \tilde K^{\rm M}_{\xi^i \xi^j} -  \tilde K_{\xi^i L^\alpha}^{\rm M} \tilde K^{{\rm M}\ L^\alpha L^\beta} \tilde K_{L^\beta \xi^j }^{\rm M}  \big)\big|_*\nn  \\
  &=&  - R \cdot K^{\rm M}_{\xi^i \xi^j}\big|_* \ ,
\eea
where one has to apply \eqref{usefull1} and \eqref{firstvanish} to evaluate the last two equalities.
$f_{ij}$ is the holomorphic gauge-coupling function of the vectors $A^i_4$. 

In general, the holomorphic gauge coupling function $f_{ij}$ can be of the form 
\beq \label{holom_fij_MF}
   \R\, f_{ij} =  \R (\cC^\alpha_{ij} \, T_\alpha - \tilde f_{ij}(M) + \cO(e^{-T}))\ ,    
\eeq
where $\cC^\alpha_{ij}$ are real constants and $\tilde f_{ij}(M)$ is a homomorphic function 
in the remaining complex scalars $M^I$. Note that the perturbative shift symmetry for
$T_\alpha$ in three dimensions prevents $T_\alpha$ to appear with additional perturbative 
contributions. This can be traced back to the fact that $T_\alpha$ arises by 
dualizing a three-dimensional vector multiplet $(A^\alpha,v^\alpha)$. We next evaluate \eqref{gauge_coupling1} and 
\eqref{holom_fij_MF}, neglecting $T_\alpha$ instanton corrections. Using the 
expression \eqref{Talpha_Leg1} one obtains the differential equation 
\beq \label{diff_eq}
    \big(\cC^\alpha_{ij} \partial_{L^\alpha}  +  R \partial_{\xi^i} \partial_{\xi^j} \big)\tilde K^{\rm M} 
    =  \R\, \tilde f_{ij}(M)  \ , \qquad \text{for} \ R,\zeta^i \rightarrow 0\ .  
\eeq
This equation poses constraints on the M-theory kinetic potential to ensure that this 
three-dimensional theory can be obtained by dimensional reduction from a 
four-dimensional effective theory with holomorphic gauge-coupling function $f_{ij}$. It 
is not hard to check that $\tilde K^{\rm M}$ as given in \eqref{tildeKM1} satisfies \eqref{diff_eq}
for $C^{S'}_{ij} =  C_{ij}$ with all other $C^\alpha_{ij}$ vanishing and $\tilde f_{ij}=0$. 

In the limit  \eqref{F-theory-lift2} the $A_i^3,\xi^i$ 
are expected to combine into a non-Abelian potential $F_4 = dA_4 + [A_4,A_4]$ for the group $G$. Here the missing 
degrees of freedom appear due to M2-branes wrapped on the vanishing $\bbP^1$'s in the fibers of the $\hat D_i$. 
This implies that 
\beq
   \int_{\mathbb{M}_{2,1}} \R f_{ij}\, F^i \wedge *_3 F^j\quad \rightarrow \quad \int_{\mathbb{M}_{3,1}} \R f_{G}\, \text{Tr} (F_{4} \wedge *_4 F_4)\ ,
\eeq
where $f_G$ is the holomorphic gauge coupling function of the four-dimensional non-Abelian gauge theory. 
Let us now apply the F-theory lift, to the large volume expression \eqref{tildeKM1} for $\tilde K^{\rm M}$.
Performing the Legendre transform and a $R,\xi^i$ Taylor expansion one finds that $\mathbf{K}^{\rm M}$ is given by 
\beq \label{explicitbfK}
   \mathbf{K}^{\rm M} = \log R + \log[ \cV_{L} (T+\bar T)] + K^{\rm cs}(z) - \frac{1}{2 R}\, \R T_{S'}\, C_{ij} \xi^i \xi^j +...\ .  
\eeq
where we have to use \eqref{special_Talpha} to evaluate $L^\alpha(T)$ as a function of $T_\alpha+\bar T_\alpha$.
The term proportional to $R^{-1}$ in this expansion is directly evaluated by using \eqref{gauge_coupling1} 
and \eqref{KijM}. Comparing \eqref{explicitbfK} to the general expression \eqref{KM_claim2} one 
easily determines the four-dimensional K\"ahler potential $K^{\rm F}$ and gauge coupling function $f_{ij}$,
\bea \label{TS'}
   K^{\rm F}(z,T) &=& \log  \cV_{L} (T+\bar T) +K^{\rm cs}(z)  \nn \\
         &=& - 2 \log \cV_{\rm b} - \log \int_{\hat X_4} \Omega \wedge \bar \Omega \ , \\ 
   \qquad f_{G}(T) &=& T_{S'}\ , \nn
\eea
where we inserted \eqref{def-Kcs_gen}, and $L^\alpha_b = v^\alpha_{\rm b}/\cV_{\rm b}$ after using \eqref{F-theory-lift2}.
Note that this precisely agrees with the expectation for a seven-brane wrapped on the 
divisor $\cS$ in the base $B_3$ \cite{Jockers:2004yj,Beasley:2008dc}.

\section{Gauge theories from the R-R sector} \label{RRsection}

In this section we discuss the gauge theory arising   
from vector fields obtained by the reduction of the 
R-R four-form $C_4$ in an F-theory compactification
on a Calabi-Yau fourfold with $2 r=2 h^{2,1}(B_3)$ harmonic  
three-forms. We begin to review some general facts about the four-dimensional 
$U(1)^r$ gauge theory in subsection \ref{IIBperspective}. The F-theory gauge couplings  
are determined by lifting a three-dimensional M-theory compactifications on $X_4$ 
to four dimensions in subsection \ref{F-dual_KK}. In the three-dimensional theory the 
vector fields are dual to complex scalars. In subsection \ref{em_duality} we comment on the 
action of four-dimensional electro-magnetic duality on the three-dimensional effective 
theory constraining the form of the superpotential. 

\subsection{Type IIB perspective and the four-dimensional action \label{IIBperspective}}

Let us first recall some facts about the four-dimensional gauge theory arising 
from the massless vector modes of $C_4$. Explicitly the gauge fields arise in the expansion
\beq \label{C_4exp}
  C_4|_{vec} = V^\kappa \wedge \alpha_\kappa - \tilde V_\kappa \wedge \beta^\kappa \ , \qquad \kappa = 1,\ldots, r\ ,
\eeq 
where we have displayed the `electric' and `magnetic' four-dimensional vectors $V^\kappa, \tilde V_\kappa$. The real symplectic basis $(\alpha_\kappa,\beta^\kappa)$
of $H^{3}(B_3,\bbZ)$ obeys
\beq \label{symplectic_pairing}
  \int_{B_3} \alpha_\lambda \wedge \beta^\kappa  = \delta^\kappa_\lambda\ , \qquad \int_{B_3} \alpha_\kappa \wedge \alpha_\lambda = \int_{B_3} \beta^\kappa \wedge \beta^\lambda = 0 \ .
\eeq
Note that $C_4$ in Type IIB supergravity 
has a self-dual five-form field strength $F_5 = *_{10} F_5$ and hence only half of the vectors in \eqref{C_4exp}
parametrize independent degrees of freedom. As we argue in the 
next  subsection \ref{F-dual_KK} one expects that this self-duality is generalized in 
an F-theory compactification. Despite this generalization the choices for splitting the gauge 
fields into sets $V^\kappa$ and $\tilde V_\kappa$ will be related 
by symplectic rotations in $Sp(2 r, \bbZ)$ of the basis $(\alpha_\kappa,\beta^\kappa)$ preserving \eqref{symplectic_pairing}. On the level of the four-dimensional effective action these 
rotations correspond to electro-magnetic rotations as we recall next.

Let us summarize some general facts about electro-magnetic duality rotations. 
It is well-known that the 
four-dimensional $\cN=1$ action for $U(1)$ gauge fields is of the form 
\beq \label{S_U(1)}
 S_{U(1)}^4 = - \int_{\mathbb{M}_{3,1}} \tfrac{1}{2} \R f_{\kappa \lambda}\,  \cF^\kappa \wedge *_4 \cF^\lambda 
              + \tfrac12 \I f_{\kappa \lambda}\, \cF^\kappa \wedge  \cF^\lambda
\eeq
where $\cF^\kappa = dV^\kappa$, and $f_{\kappa \lambda}$ is a holomorphic function of the 
complex scalars in the chiral multiplets. The electro-magnetic rotations mix Bianchi identities and equations of motion for the 
gauge fields $V^\kappa$. The elements of the symplectic group satisfy
\beq \label{symp_rot}
    \left( \begin{array}{cc}A & B \\ C & D \end{array} \right) \in Sp(2 r, \bbZ)\ , \qquad  
    \begin{array}{c}D^T A - B^T C = 1\ ,  \\[.1cm] A^T C = C^T A, \ B^T D = D^T B\ ,  \end{array} 
\eeq
where $A,B,C,D$ are $r\times r$ matrices.
The matrix \eqref{symp_rot} acts linearly on the vector $(\cG_\kappa,\cF^\kappa)$, where 
\beq  \label{def-Gkappa}
  \cG_\kappa = -\delta S_{U(1)}^4 / \delta \cF^\kappa =  \R f_{\kappa \lambda}\,  *_4 \cF^\lambda + \I f_{\kappa \lambda}\,  \cF^\lambda\ . 
\eeq 
Note that the gauge fields couple to the complex scalars of the theory via the holomorphic 
function $f_{\kappa \lambda}$. Hence, a electro-magnetic rotation can be induced 
by a transformation of the scalars and has to be accompanied with a rotation of the 
coupling matrix $\mathbf{f}$  with entries $\mathbf{f}_{\kappa \lambda}\equiv -i f_{\kappa \lambda}$ as
\beq \label{rot_f}
  \mathbf{f} \ \rightarrow \ (A\mathbf{f}+B) (C\mathbf{f}+D)^{-1} \ . 
\eeq
The factor $-i$ arises from the convention that $\R f_{\kappa \lambda} = (1/g^2)_{\kappa \lambda}$ is 
positive definite since it defines the inverse gauge-coupling.

In Type IIB supergravity $\cG_\kappa$ given in \eqref{def-Gkappa} admits an identification as the field strength of $\tilde V_\kappa$.
However, in a general F-theory compactification it is not expected that this can be inferred 
from the classical self-duality of $F_5$, since $f_{\kappa \lambda}$ will in general 
also depend on other moduli, such as the deformations of the seven-branes.
We determine the $f_{\kappa \lambda}$ in terms of the geometry of the 
Calabi-Yau fourfold in the next subsection.

\subsection{Bulk gauge theory in the F-theory lift}\label{F-dual_KK}

In order to study the F-theory dynamics and couplings of the vectors in \eqref{C_4exp} in 
more detail we again have to perform an M- to F-theory lift. In order to simplify the 
discussion, we restrict ourselves in this section to Calabi-Yau fourfolds with  
\beq \label{only_vec}
   r \equiv h^{2,1}(\hat X_4) = h^{2,1}(B_3)\ .
\eeq
Moreover, recall that $h^{3,0}(\hat X_4)=0$ such that the whole third cohomology 
splits as 
\beq
   H^{3}(\hat X_4,\bbC) = H^{2,1}(\hat X_4)\oplus H^{1,2}(\hat X_4)= H^{2,1}(B_3)\oplus H^{1,2}(B_3)\ .
\eeq
This split depends on the complex structure on $\hat X_4$ and hence varies non-trivially over 
the complex structure moduli space $\cM^{\rm cs}$ discussed in section \ref{complex_structure_section_1}. 
Moreover, in the quantum theory one actually has 
to consider the torus bundle $\mathbb{T} \rightarrow \cM^{\rm cs}$ with complex $r$-torus 
fibers
\beq \label{torus-fiber}
   \mathbb{T}_z = H^{2,1}(\hat X_4)/H^3(\hat X_4,\mathbb{Z}) \cong H^{2,1}(B_3)/H^3(B_3,\mathbb{Z})\ ,
\eeq 
and base $\cM^{\rm cs}$. At special points in $\cM^{\rm cs}$ the fibers 
$\mathbb{T}_z$ can become singular signaling that the effective theory was not properly 
determined since light degrees of freedom have been improperly discarded. 
At generic points in the moduli space one finds 
an $\cN=1$, $U(1)^r$ gauge theory after the lift of the three-dimensional M-theory 
compactification to a four-dimensional F-theory compactification.
Note 
that the study of such torus fibration has recently attracted much attention 
in the context of reductions of $\cN=2$ gauge theories from four to three dimensions 
(see refs.~\cite{Gaiotto:2008cd} for recent progress and further references). Clearly, our 
$\cN=1$ set-up is much less constraint by supersymmetry. While in $\cN=2$ the 
torus bundle is of dimension $4r$ and admits a Hyperk\"ahler metric
the $\cN=1$ torus bundle $\mathbb{T}$ is of dimension 
$2r+2 h^{3,1}(\hat X_4)$ and has a K\"ahler metric. One of the tasks of this section is 
to determine the classical form of this metric in a theory coupled to gravity.

As already noted in section \ref{spectrum_summary} imposing \eqref{only_vec} ensures that all non-trivial three-forms on $\hat X_4$ 
descend to four-dimensional vector multiplets in the F-theory limit. In the M-theory reduction these
vectors arise from expanding the 
M-theory three-form $C_3$ into $(\alpha_\kappa,\beta^\kappa)$ via the Kaluza-Klein Ansatz
\bea \label{C3expansion_3}
  C_3 &=& A^\cA\wedge e_\cA + \tilde a^\kappa\, \alpha_\kappa -\tilde b_\kappa \, \beta^\kappa \nn \\
      &=& A^\cA \wedge e_\cA + \cN_\kappa\, \psi^\kappa + \bar \cN_\kappa\, \bar \psi^{\kappa}\ ,   
\eea
where $A^\cA=(A^0,A^i,A^\alpha)$ are vectors which we already included in \eqref{C3expansion_1}, 
and $(\tilde a^\kappa,\tilde b_\kappa)$, $\cN_\kappa$ are real and complex scalars in three dimensions.
In three-dimensions massless scalars are dual to vectors and we will show momentarily, that after 
dualizing $\tilde b_\kappa$ into a vector $V^\kappa_3$, the fields $(V_3^\kappa,a^\kappa)$ comprise the degrees of freedom of a four-dimensional vector 
$V^\kappa$ in \eqref{C_4exp}.

In \eqref{C3expansion_3} we have also introduced a complex basis of $(2,1)$-forms $\psi^\kappa$ of $H^{2,1}(\hat X_4)$.
The basis elements $\psi^\kappa$ depend on the complex structure of $\hat X_4$ and naturally define 
complex coordinates $\cN_{\kappa}$.
Let us explore the relation between the complex and real basis for $H^3(\hat X_4)$.
In general, one can identify 
\beq \label{psi_exp}
  \psi^\kappa =  \tfrac{1}{2} \R f^{\kappa \lambda} (\alpha_\lambda  - i   \bar f_{\lambda \mu} \beta^\mu) \ , \qquad \psi^\kappa - \bar \psi^\kappa = -i \beta^\kappa\ ,
\eeq
for a complex function $f_{\kappa \lambda}$ of the complex structure moduli $z^\cK$, with $\R f^{\kappa \lambda}\equiv (\R f_{\kappa \lambda})^{-1}$
being the inverse of the real part of $f_{\kappa \lambda}$. One can now show that for an appropriate 
choice of $\psi^\kappa$ the function $f_{\kappa \lambda}(z)$ is holomorphic 
in $z^\cK$. This can be deduced from the fact  
that for a complex manifold $\hat X_4$ the filtration $F^3(\hat X_4) = H^{3,0}$,  
$F^2(\hat X_4) = H^{3,0}  \oplus H^{2,1}$, 
etc.\ consists of holomorphic bundles $F^i(\hat X_4)$ over the space of complex 
structure deformations~\cite{Voisin}. Since $H^{3,0}$ is trivial one finds that 
$F^2=H^{2,1}$ is a holomorphic bundle and one can locally choose a basis
$\psi^\kappa(z)$ as in \eqref{psi_exp}. The matrix $f_{\kappa \lambda}$ is 
readily extracted from $\psi^\kappa$ using \eqref{symplectic_pairing}.
As an immediate consequence of \eqref{C3expansion_3} and \eqref{psi_exp} one concludes that 
\beq \label{defNkappa}
  \cN_\kappa =  f_{\kappa \lambda}(z)\, \tilde a^\lambda - i \tilde b_\kappa = - i (\tilde b_\kappa - \mathbf{f}_{\kappa \lambda} \tilde a^\lambda)\ ,
\eeq
where $ \mathbf{f}_{\kappa \lambda} = - i f_{\kappa \lambda}$ as in \eqref{rot_f}.
These expressions, together with the analysis of the couplings of $\cN_\kappa$, 
allows to identify the $\cN_\kappa$ as arising from four-dimensional vectors $V^\kappa$
after reduction to three dimensions on a circle. Moreover, the function $f_{\kappa \lambda}(z)$
is the holomorphic four-dimensional gauge coupling function in \eqref{S_U(1)}.
The reduction from four to three dimensions is reviewed in appendix \ref{scalar_vec_duality}. To fully 
justify this identification under reduction one also has to analyze the 
couplings of the scalars $\cN_\kappa$, and hence derive the three-dimensional kinetic 
potential including the $\cN_\kappa$.

To identify the appearance of the moduli 
$\cN_\kappa$ in the M-theory kinetic potential one 
dimensionally reduces the eleven-dimensional kinetic term for $G_4=dC_3$ and the 
Chern-Simons term 
\bea \label{CSred_1}
   S^{11}_{G_4}&=& -\int \tfrac{1}{4} G_4 \wedge *G_4 + \tfrac{1}{12} C_3 \wedge G_4 \wedge G_4  \\
   &=& - \int_{\mathbb{M}_{2,1}} \cG^{\kappa \lambda} D \cN_\kappa \wedge * \overline{D \cN_\lambda} 
- \tfrac{1}{2} d_{\cA}^{\phantom{\cA} \kappa \bar \lambda}\, F^\cA \wedge \I(\bar \cN_\kappa {D \cN}_\lambda)+\ldots \ , \nn
\eea
where $D \cN_\kappa = d \cN_\kappa -  \R \cN_\lambda \,   \R f^{\lambda \mu}\, d f_{\mu \kappa}$.
The metric on the space of three-forms is given by 
\beq \label{d_AIJ}
  \cG^{\kappa \lambda} = \frac{1}{2 \cV} \int_{\hat X_4} \psi^\kappa \wedge * \bar \psi^{\lambda}= - \frac{v^\cA d_{\cA}^{\phantom{\cA} \kappa \bar \lambda}}{2 \cV}\ ,
  \qquad d_{\cA}^{\phantom{\cA} \kappa \bar \lambda} = i\int_{\hat X_4} \omega_\cA \wedge \psi^\kappa \wedge \bar \psi^\lambda\ ,
\eeq 
where we have used $*\bar \psi^\kappa = -iJ\wedge \bar \psi^\kappa$.
The coupling $d_{\cA}^{\phantom{\cA} \kappa \bar \lambda}$ depends on 
the complex structure moduli through the complex three-forms $\psi^\kappa$. 
The crucial point to note is that \eqref{CSred_1} also induces a coupling 
of $\cN_\kappa$ and the complex structure moduli $z^\cK$ to the three-dimensional vectors $A^\cA$. 
Note that many of the above statements, in particular equations \eqref{psi_exp} and \eqref{CSred_1}, 
are independent of the restriction \eqref{only_vec}. 
However, in case we restrict to geometries where all non-trivial three-forms arise from the 
base $B_3$, i.e.~\eqref{only_vec} is obeyed, one can further 
deduce that the only non-vanishing $d_{\cA}^{\phantom{\cA} \kappa \bar \lambda} $ is 
along $\omega_0$, the  two-form Poincar\'e dual to $B_3$ in $\hat X_4$. Explicitly, one has  
\beq
  \cG^{\kappa \lambda} = - \tfrac{1}{2} R \cdot d_{0}^{\phantom{\cA} \kappa \bar \lambda}\ , \qquad d_{0}^{\phantom{\cA} \kappa \bar \lambda} = i \int_{B_3}  \psi^\kappa \wedge \bar \psi^\lambda = - \tfrac12 \R f^{\kappa \lambda}\ ,
\eeq
where we inserted \eqref{psi_exp}. By comparing  
\eqref{CSred_1} with \eqref{kinetic_lin_gen_1} one thus infers 
that $\cN_\kappa$ appears in the kinetic potential $ \mathbf{K}^{\rm M}$
as 
\beq \label{correctedtildeK_vec}
  \mathbf{K}^{\rm M} = \log R + K^F(z,T) -\frac{1}{2R} \R T_{S'} C_{ij} \xi^i \xi^j + \frac12 R\cdot \R f^{\kappa \lambda} (z)\, \R \cN_\kappa \, \R \cN_\lambda\ ,
\eeq
where $K^F(z,T)$ is the $\cN_\kappa$-independent four-dimensional K\"ahler potential \eqref{TS'}. 
Note that this kinetic potential reproduces correctly the first term in the reduction of \eqref{CSred_1}. 
The second term in the reduction \eqref{CSred_1} is only reproduced up to a total derivative.
It is now readily checked that \eqref{correctedtildeK_vec} indeed 
encodes the dynamics of four-dimensional vectors $V^\kappa$. Comparing 
\eqref{correctedtildeK_vec} to the general expression 
\eqref{K1_exp}, obtained by dimensional reduction of the general $\cN=1$ four-dimensional action, one 
finds 
\beq
  f_{G} = T_{S'} \ , \qquad \qquad f^{\rm RR}_{\kappa \lambda} = f_{\kappa \lambda}(z)\ . 
\eeq 
where $f^{\rm RR}_{\kappa \lambda}$ is the four-dimensional gauge coupling function 
of the R-R $U(1)$'s.

\subsection{Electro-magnetic duality in the three-dimensional action} \label{em_duality}

Having established the couplings of the $\cN_\kappa$ in the three-dimensional action, it is interesting 
to investigate the action of the symplectic group \eqref{symp_rot} when acting linearly
on the basis vector $(\alpha_\kappa,\beta^\kappa)$ and $(\tilde b_\kappa,\tilde a^\kappa)$ in \eqref{C3expansion_3}. To explore 
that in more detail, we compute
\beq \label{red-T0}
  T_0 = \partial_R  \mathbf{K}^{\rm M}  + i c_0 
      = \tilde T_0 + \tfrac14 \R f^{\kappa \lambda} \cN_\kappa (\cN_\lambda +\bar \cN_\lambda)\ ,
\eeq
where $\tilde T_0$ is invariant under the transformations \eqref{symp_rot}.\footnote{Note that 
as in \cite{Grimm:2007xm} we have shifted the imaginary part $c_0$ to $\I \tilde T_0 = c_0 - \frac12 \R f^{\kappa \lambda} \I \cN_\kappa \R \cN_\lambda$
claiming that this $\tilde T_0$ is the correct invariant combination.}
It is not hard to check that $\mathbf{f}_{\kappa \lambda} = - i f_{\kappa \lambda}$ transforms 
as in  \eqref{rot_f}. 
We evaluate the transformations of $\cN_\kappa,T_0$ under \eqref{symp_rot} by using the invariance of $\tilde T_0$ as
\beq \label{act_symp_rot}
  \cN_\kappa  \rightarrow (C \mathbf{f} + D)^{-1 \lambda}_{\kappa} \cN_\lambda \ ,\qquad T_0  \ \rightarrow \  T_0 + \tfrac{i}{2} C^{\mu \kappa} (C \mathbf{f} +D)^{-1 \lambda}_{\kappa} \cN_\lambda \cN_\mu \ .
\eeq
As a further set of transformations one can evaluate the behavior under 
integral shifts $n^\kappa$ of $\tilde a^\kappa$. One finds that
\beq \label{act_shift}
  \cN_\kappa \ \rightarrow \ \cN_\kappa + 2\pi i\, \mathbf{f}_{\kappa \lambda} n^\kappa\ ,\qquad 
  T_0  \ \rightarrow \  T_0 - 2\pi n^\kappa \cN_\kappa - 2 \pi^2 i\, \mathbf{f}_{\kappa \lambda} n^\kappa n^\lambda \ .\ .
\eeq
Note that the non-trivial shifts \eqref{act_symp_rot} and \eqref{act_shift} for $T_0$ are
expected from an analysis of the M5-brane action \cite{Witten:1996hc,Ganor:1996pe}.

The transformation properties \eqref{rot_f}, \eqref{act_symp_rot} and \eqref{act_shift} 
can be used to constrain the couplings. In particular, a subgroup $H\subset Sp(2r,\bbZ)$ might provide an actual 
symmetry group of the four-dimensional gauge theory. This group has to be determined 
by studying the monodromies of the torus fibration \eqref{torus-fiber} over the 
complex structure moduli space of $\hat X_4$. 
In other words, since $f_{\kappa \lambda}$ 
depends on the complex structure deformations of $\hat X_4$, a symmetry of $\cM^{\rm cs}$ can 
induce an $H$ action on the gauge-fields. Of particular interest are the monodromy 
symmetries \eqref{monodromy_group} of $\cM^{\rm cs}$. Using intuition from the orientifold limes 
one expects that there exist Calabi-Yau fourfolds with a natural lattice embedding 
\beq
   H^{3}(\hat X_4,\bbZ) \ \hookrightarrow \ H^{4}(\hat X_4, \bbZ)\ .
\eeq 
This implies an action of the monodromy group \eqref{monodromy_group} of $\cM^{\rm cs}$ on 
the gauge fields from the R-R sector such that $H \subset G^{\rm sym}$. This would be reminiscent of an 
underlying $\cN=2$ theory. However, in contrast to the $\cN=2$ theory 
the K\"ahler moduli sector discussed in sections \ref{Fsystematics} and \ref{non_Abelian} can 
directly correct the gauge coupling functions in this $\cN=1$ setting.
Nevertheless, as we show in a forthcoming publication, 
it is interesting to explore the link between the geometry 
of the complex structure moduli space $\cM^{\rm cs}$ and the 
$\cN=1$ gauge theory of this section.

One can now proceed as in refs.~\cite{Witten:1996hc,Ganor:1996pe,Grimm:2007xm} and constrain the form of the last term 
in three-dimensional superpotential  \eqref{WM_exp} as
\beq \label{WT0}
   W^{\rm M}(z,\cN,T_0) = \tilde \cC\cdot \Theta(\mathbf{f},\cN) e^{-T_0} \ ,
\eeq
where $\tilde C$ can still be a holomorphic function in $z^\cK$. The key point to note is 
that $\Theta(\mathbf{f},\cN)$ transforms under \eqref{symp_rot} and the integral 
shifts of $(a^\kappa,\tilde b_\kappa)$ such that it cancels the shifts of $T_0$. This allows to 
identify $\Theta(\mathbf{f},\cN)$ to be a Jacobi form \cite{EZ}. It is an interesting 
task to determine explicitly the form of  $\Theta(\mathbf{f},\cN)$ for a given Calabi-Yau fourfold with $h^{2,1}(B_3)>0$.
Moreover, it would be interesting to investigate a four-dimensional interpretation of \eqref{WT0}, by 
recalling that the leading term in $T_0$ is the action of a Taub-NUT gravitational 
instanton \cite{DGK}.\footnote{See refs.~\cite{Blumenhagen:2010ja,Cveti{c}:2010rq}, for recent progress on F-theory instantons.}
 
Let us end this section by giving one simple example of a Calabi-Yau fourfold in which the above formalism can be applied. 
Namely, one can consider the elliptic fibration over the cubic hypersurface $B_3$ in $\bbP^4$. The cubic is a 
Fano threefold with $h^{2,1}(B_3)=5,\, h^{1,1}(B_3)=1$. The corresponding elliptically fibered Calabi-Yau fourfold $X_4$ 
is constructed as complete intersection as summarized in appendix A of ref.~\cite{Grimm:2009yu}. Its Hodge data are
\beq
   h^{1,1}(X_4) = 2\ , \quad h^{2,1}(X_4) = 5 \ ,\quad h^{3,1} = 1483\ ,
\eeq
where the two $(1,1)$-forms correspond to the hyperplane class in $B_3$ and the fiber of $X_4$. Note that $X_4$ is 
generically non-singular, so that no non-Abelian gauge symmetry arises from space-time filling 
seven-branes.\footnote{By tuning the complex structure non-trivial gauge-theories arise as 
discussed in \cite{Grimm:2009yu}.} The cubic threefold and its intermediate Jacobian $H^{2,1}(B_3)/H^{3}(B_3,\bbZ)$
has been studied in detail in ref.~\cite{Clemens}. In this case only the dependence on the complex structure 
moduli of $B_3$ was included. It would be interesting to extend this analysis to the whole fourfold $X_4$.
Moreover, such an analysis is particularly interesting in the case that the 
R-R gauge theory can be traced through heterotic F-theory duality. It was argued 
in refs.~\cite{Lukas:1998hk,Berglund:1998ej,Rajesh:1998ik} that the $U(1)$'s 
in F-theory arise in the heterotic dual as gauge fields on heterotic M5-branes wrapped 
on a curve $\cC$ in the heterotic compactification manifold.\footnote{See refs.~\cite{Grimm:2009sy,Jockers:2009ti} for 
recent discussions on this duality.} 
We hope to report on progress along these lines in a future publication.

\section{On matter couplings on seven-branes} \label{matter_couplings}

In this section we discuss how certain matter couplings on seven-branes 
are encoded in the $\cN=1$ F-theory effective action. In subsection \ref{world-volume-theory}
we review briefly aspects of the local seven-brane world volume theory. 
Adjoint matter encoding deformations of the seven-branes is discussed in subsection 
\ref{Adjoint_deforms}, where we also present a detailed determination the D-term supersymmetry 
conditions. Adjoint Wilson line moduli and scalars from the Type IIB 
R-R and NS-NS two-forms are discussed in subsection \ref{Wilson}.

\subsection{Seven-brane world volume theory} \label{world-volume-theory}

In a local analysis of the world-volume theory on a stack of seven-branes 
wrapped on $\cS$ it was shown in ref.~\cite{Cecotti:2009zf,Donagi:2008ca,Beasley:2008dc} that the zero modes can be obtained by 
solving eight-dimensional F- and D-term equations. 

To make this more explicit, we first specify a background seven-brane configuration extracted 
from $X_4^{\rm sing}$ or $\hat X_4$. Recall from section \ref{Ell+seven} 
that the local gauge group at a point $p$ in $\cS$ is determined by 
the ADE degeneration of the elliptic fibration $X_4^{\rm sing}$ at $p$. Let us denote 
by $G^{\rm max}$ the maximal local gauge group which appears when considering all points on $\cS$. 
We will consider in the following cases where all other points of $\cS$ have 
local gauge groups inside $G^{\rm max}$. The actual physical gauge group $G$ is obtained 
at generic points in $\cS$ where no further enhancement takes place. Further enhancements 
will arise over complex \textit{matter curves} 
obtained at the intersection of $\cS$ with 
the locus $\Delta'$ given in \eqref{Delta_split}, as well as \textit{Yukawa points} on $\cS$ where matter curves meet. 
This information can be explicitly extracted for a given $X_4^{\rm sing}$, or $\hat X_4$. To specify the 
background of the local field 
theory on the seven-brane, this data is conveniently encoded by a $(2,0)$ form $\langle \varphi \rangle$ 
on $\cS$. In the simplest case, when the breaking of $G^{\rm max}$ over the matter curves 
can be captured by a vev in the Cartan subalgebra $\mathfrak{h}_{{\rm max}}$ to $G^{\rm max}$ one has
\beq
   \langle \varphi \rangle \ \in\ \mathfrak{h}_{{\rm max}} \otimes K \cS\ , 
\eeq
which is specified for a fixed complex structure of $X^{\rm sing}_4$. Each non-trivial 
entry of $\langle \varphi \rangle$ specifies a breaking of $G^{\rm max}$. Due to 
a minimal gauge group $G$ on $\cS$ one has at least $rk(G)$ vanishing 
entries in the background configuration.
Moreover, note that over the matter curves additional vanishing conditions 
apply, which ensure that the gauge group locally enhances further. 
Further data specifying the background are given by 
$\langle A \rangle$, the background value of the gauge field $A$ on $\cS$.
The four-dimensional effective theory is computed for the 
fluctuations $\varphi'$ and $A'$ around such a background configuration, i.e.~one 
expands 
\beq
   \varphi =\langle \varphi \rangle + \varphi' \ , \qquad A = \langle A \rangle + A'\ .
\eeq 
At low energies only the zero-modes for $\varphi',A'$ will appear. These are determined 
by solving the eight-dimensional F- and D-term equation expanded around the 
background as we recall next.
  
Let us denote by $\cA$ the $(1,0)$-part of the seven-brane gauge field $A$ in 
a fixed complex structure of $X^{\rm sing}_4$.
The eight-dimensional F- and D-term equations in this 
complex structure are given by
\bea \label{8dF-term}
  \text{8d F-term}: &\ & \bar \partial_{\bar \cA} \varphi = 0 \ , \qquad F^{0,2} = 0 \ ,\\
  \text{8d D-term}: &\ & \omega \wedge F + \tfrac{i}{2}[\bar \varphi,\varphi] = 0 \ , \label{8dDterm}
\eea
where $\omega = c J_{\rm b}|_{\cS}$ with $c$ constant on $\cS$ as determined below, $F = dA + A \wedge A$, and $\partial_\cA = \partial + \cA \wedge $ is the gauge-covariant 
derivative on $\mathbb{M}_{3,1}\times \cS$. Note that the F-term equations admit a gauge invariance 
\beq \label{A-gauge}
   \bar \cA \ \rightarrow\ g^{-1} \bar \cA g + g^{-1} \bar \partial g\ , \qquad \varphi \ \rightarrow \ g^{-1} \varphi g\ .
\eeq

In order to extract the light modes appearing in the four-dimensional effective 
action one first expands the F- and D-term equations to linear order in the 
fluctuations $\varphi',A'$ as 
\bea \label{zero_modes}
   &&\bar \partial_{\langle A\rangle} \bar \cA' = 0\ ,\qquad  
   \bar \partial_{\langle A\rangle} \varphi'+[\bar \cA',\langle \varphi \rangle] = 0\ , \\
   && \omega \wedge \partial_{\langle A\rangle} \bar \cA' + \tfrac{i}{2} [\langle \bar \varphi \rangle, \varphi'] = 0\ . 
   \label{lin_D}
\eea
It was argued in ref.~\cite{Cecotti:2009zf} that finding the zero modes for the F-term equations 
\eqref{zero_modes}, can be studied by determining the cohomology to a particular differential operator.
Each class is independent of the K\"ahler form due to the gauge invariance \eqref{A-gauge}.
The linearized D-term conditions \eqref{lin_D} then determine a specific representative in a cohomology class. 
Of course, that is similar to, for example, the standard Kaluza-Klein Ansatz for the M-theory three-form \eqref{C3-exp_simple}, \eqref{C3expansion_1} and \eqref{C3expansion_3}, where the expansion forms are representing cohomology classes and are independent of the 
K\"ahler moduli. Despite such a cohomology theory 
it is not immediately clear how to include the zero modes $A',\varphi'$ into 
a Kaluza-Klein reduction. One of the complications is the coupling of $\varphi'$ and 
$A'$ zero modes in the second equation of \eqref{zero_modes}. 

In the following we will restrict to the simplest situation and consider the special 
case of matter $\varphi'$ and $A'$ transforming in the adjoint of $G$, and set $\langle A \rangle=0$.
Since the background $\langle \varphi \rangle$ has zero entries along the adjoint of $G$ on $\cS$, the 
conditions \eqref{zero_modes} reduce to $\bar \partial \bar \cA' = 0$ and $\bar \partial \varphi' = 0$.
These conditions are satisfied if $\bar \cA',\varphi'$ are expanded in a basis of 
$H^{0,1}(\cS)$ and $H^{2,0}(\cS)$ with coefficients transforming 
in the adjoint representation. In the remaining two subsections we 
study this Kaluza-Klein Ansatz in an F- and M-theory reduction.

\subsection{Adjoint matter on the seven-brane world volume} \label{Adjoint_deforms}

In this subsection we discuss the inclusion of adjoint matter $\varphi'$ 
localized on the world volume of a seven-brane with gauge group $G$.  In the 
four-dimensional effective theory $\varphi'$ is expanded into zero modes $\rho_\nu$ forming a basis of $H^{2,0}(\cS)$, with 
coefficients being the four-dimensional matter fields $\varphi'^\nu$.
The degrees of freedom captured by $\varphi$ can be turned into additional 
complex structure deformations if one moves from the resolved Calabi-Yau fourfold $\hat X_4$ 
to the deformed Calabi-Yau fourfold $X_4$ as in \eqref{res_deform}. This suggests that at leading 
order the deformations captured by $\varphi'$ appear in the K\"ahler potential as
\beq \label{K_varphi}
  K^{\rm cs}(z,\bar z,\varphi',\bar \varphi') = - \log \big[ \int_{\hat X_4} \Omega \wedge \bar \Omega - \int_{\cS} \text{Tr}(\varphi' \wedge \bar \varphi') \big]\ ,
\eeq
where the trace is in the adjoint of $G$. In the following 
we compute the four-dimensional D-term induced via the 
minimal coupling to adjoint matter and a non-trivial background flux.

One first notes
that there is now charged adjoint matter $\varphi'$ coupling via the covariant 
derivative $D\varphi'^\nu = d\varphi'^\nu + [A_4 ,\varphi'^\nu]$. To 
compute the D-term arising due to this gauging, we recall some general 
facts about supergravity theories with four supercharges.
Denoting by $M^I$ all complex scalars in chiral multiplets, 
the Killing vector $X_i^I(M,\bar M)$ of vectors $A^i$ appear in the minimal coupling 
$D M^I = dM^I + i X_i^I A^i $.
The D-term for the vector multiplet with $A^i$ is now evaluated using the general 
four-dimensional supergravity identity \cite{Wess:1992cp}
\beq \label{general_D}
   \partial_{M^I} D_i =  K_{I \bar J} \bar X_{i}^{\bar J}\ ,\qquad \quad K_{I \bar J} = \partial_{M^I}\partial_{\bar M^J} K\ .
\eeq
One thus has to evaluate the Killing vector $X_{A_4}^{\varphi^\nu}$ and the derivative of the K\"ahler potential 
for $\varphi'^\nu$. Using 
\eqref{K_varphi} one finds at leading order
\beq
X_{A_4}^{\varphi^\nu} = - i[\cdot\, ,\varphi'^{\nu}]\ ,\qquad \quad
   \partial_{\varphi'^\nu} \partial_{\bar \varphi'^\mu} K^{\rm cs} = \frac{\int_{\cS} \rho_{\nu} \wedge  \bar \rho_\mu}{\int_{X_4} \Omega \wedge \bar \Omega}\ . 
\eeq
This yields the leading order D-term 
\beq \label{Dgvarphi}
  D_G^{\varphi'} = \frac{i}{\int_{X_4} \Omega \wedge \bar \Omega}\int_{\cS}  [\varphi',\bar \varphi']\ .   
\eeq

Let us next include the $G_4$ flux. 
The scalar identity \eqref{general_D} reduces trivially to three space-time dimensions 
and can be directly evaluated for the M-theory reduction \cite{Haack:1999zv,Berg:2002es}. 
One focuses on the following 
terms in the reduction of eleven-dimensional supergravity action
\beq \label{CSred2}
   S_{CS} = -\frac{1}{12}\int C_3 \wedge G_4 \wedge G_4 = 
   - \frac12 \int_{\mathbb{M}_{2,1}} A^i \wedge  dA^\alpha \int_{\hat X_4} \omega_\alpha \wedge \tw_i \wedge G_4 +\ldots \ ,
\eeq
where we inserted the expansion \eqref{C3expansion_1} for $C_3$ and one of the $G_4=dC_3$. The integral 
over $\hat X_4$ arises due to the non-trivial flux. One realizes 
that this provides an additional coupling involving the vector multiplets $A^i$. Recall that the 
$A^i$ lift to four-dimensional vectors, while the $A^\alpha$ lift to four-dimensional two-forms 
and are dualized into scalars $\I\, T_\alpha$, via the 
Legendre transform \eqref{Talpha_Leg1}. Hence, the term \eqref{CSred2} is of St\"uckelberg type 
and induces a gauging of $T_\alpha$ via the covariant derivative 
\beq
  D T_\alpha = dT_\alpha + i X_{\alpha\, i} A^i\ ,\qquad    
  X_{\alpha\, i} = \frac12 \int_{\hat X_4} \omega_\alpha \wedge \tw_i \wedge G_4\ ,
\eeq
where $X_{\alpha i}$ is the flux-dependent Killing vector of the gauged shift symmetry.
 Since $X_{\alpha\,i}$ is independent 
of the chiral multiplets one can integrate \eqref{general_D} to $D_i = K_{T_\alpha} X_{\alpha\, i}$. Using 
$K_{T_\alpha} = - \frac12 L^\alpha$ for $K = \mathbf{K}^{\rm M}$ as given in \eqref{KM_id}, one finds
\beq \label{D_i}
  D_{i} = - \frac{1}{4}L^\alpha \int_{\hat X_4} \omega_\alpha\wedge \tw_i \wedge G_4 =   
          \frac{1}{4} C_{ij}\, L^\alpha \int_{\cS} \omega_\alpha \wedge F^{j}_{\rm flux}\ , 
\eeq
where we inserted the $G_4$ flux given in \eqref{G4lift}, used the identity \eqref{tw_identity},
and applied Poincar\'e duality.
It is worthwhile noting that the calculation of this D-term did not depend on the precise form of the 
K\"ahler potential, since $K_{T_\alpha} = - \frac12 L^\alpha$ is fixed by the 
Legendre transform \eqref{kin_F-dual1}, \eqref{Talpha_Leg1} and corrections to $\tilde K^{\rm M}$ will
alter the definition of $T_\alpha$ rather then $K_{T_\alpha}$. The D-term potential 
is then evaluated as 
\beq
   V_D =  \tfrac{1}{2} \R f^{-1\, ij} D_i D_j\ ,
\eeq
where $\R f_{ij}$ is the real part of the holomorphic 
gauge coupling function of the seven-branes on $\cS$ given in \eqref{holom_fij_MF}, \eqref{explicitbfK}. It is not hard to 
check that this potential term precisely arises from the dimensional reduction of 
the term $\int G_4 \wedge * G_4$ appearing in the eleven-dimensional supergravity theory.
In the F-theory lift, one thus combines \eqref{Dgvarphi} and the lifted version of \eqref{D_i}.
This yields the D-term for a non-Abelian group $G$ on $\cS$:
\beq
  D_{G} = \frac{1}{4 \cV_{\rm b}} \int_{\cS} J_{\rm b} \wedge F_{\rm flux} + \frac{i}{\int_{X_4} \Omega \wedge \bar \Omega}\int_{\cS}  [\varphi',\bar \varphi']\ .   
\eeq
Note that this agrees with the expression \eqref{8dDterm} obtained in ref.~\cite{Beasley:2008dc} for a 
seven-brane decoupled from gravity if one includes the pre-factors $\cV_{\rm b}$ and $\int_{X_4} \Omega \wedge \bar \Omega$.

\subsection{Wilson lines and R-R and NS-NS two-form moduli \label{Wilson}}

In this subsection we will include the degrees of freedom corresponding to Wilson line moduli 
on the seven-branes on a divisor $\cS$ as well as the moduli 
from the Type IIB R-R and NS-NS two-forms. 

The Wilson line degrees of freedom are present
if the Hodge-numbers $h^{1,0}(\cS)$ are non-zero. More generally they 
can arise for a non-trivial flux background, or be localized along matter curves 
as briefly discussed in subsection \ref{world-volume-theory}. Here we will focus on the simplest case 
where the Wilson lines transform in the adjoint of $G$ and arise as $h^{1,0}(\cS)$
zero modes. 
In the Kaluza-Klein zero mode expansion of the seven-brane 
gauge fields $A$ the complex Wilson line scalars $\tN^{b}$ appear as 
\beq \label{Ascalar}
   A|_{\rm scalar} = \cA + \bar \cA = \bar \tN^{b}\, \gamma_b + \tN^{b}\, \bar \gamma_b\ ,  
\eeq 
where $\gamma_b$ is a basis of $H^{1,0}(\cS)$. Note that 
we have used the split of the first cohomology group on $\cS$
by using the induced complex structure from $\hat X_4$. 
One can introduce a basis $(\hat \alpha_a,\hat \beta^b)$ of $H^{1}(\cS,\bbZ)$ and
write 
\beq \label{gamma_exp}
   \gamma_b = \hat \alpha_b - i \bar f_{bc} \hat \beta^c\ ,
\eeq
for $f_{cb}(z)$ being a holomorphic function in the complex structure moduli $z^\cK$ of $\hat X_4$.
Note that \eqref{Ascalar} and \eqref{gamma_exp} are consistent with the holomorphicity properties 
of the superpotential induced in the 
presence of a non-Abelian gauge theory. In was argued in refs.~\cite{Beasley:2008dc,Donagi:2008ca}, that 
the four-dimensional flux superpotential \eqref{W_GVW} 
is corrected by $W_{\rm brane} = \int_{\cS} \text{Tr} (F \wedge \varphi)$. 
Thus, the effective four-dimensional superpotential is given by 
\beq \label{full_WF}
  W^{\rm F} = \int_{\hat X_4} G_4 \wedge \Omega(z) + \int_{\cS} \text{Tr}(F_{\rm flux} \wedge \varphi') + Y_{ab\nu} \text{Tr}(\tN^a \tN^b \varphi'^\nu) \ , 
\eeq
with Yukawa couplings 
\beq \label{Yukawas}
  Y_{ab\nu}(z) = \int_{\cS} \bar \gamma_a \wedge \bar \gamma_c \wedge \rho_\nu \ .
\eeq
Inserting \eqref{gamma_exp} one sees that $Y_{ab\nu}$ is holomorphic in the 
complex structure deformations for a holomorphically varying basis $\rho_{\nu}$ of $(2,0)$
forms on $\cS$. 
By using the duality with the M-theory set-up, we 
discuss in the following how the Wilson line moduli $\tN^a$ appear in the 
K\"ahler potential.

To begin with, let us note that the base $B_3$ has $h^{1,0}(B_3)=0$, such that 
the one-forms in $H^{1,0}(\cS)$ have to be trivial in $B_3$. 
On the level of the Poincar\'e dual three-cycles 
$\cA^b_3 \in H_3(\cS)$ this implies that $\cA^a_3$ has to be trivial 
in $B_3$. Hence there exist four-chains which admit $\cA^a_3$ 
as boundaries. In the Calabi-Yau fourfold $\hat X_4$, one has to check 
whether or not the singular elliptic fibration makes the four-chains 
into five-cycles in $H_{5}(\hat X_4)$. This can happen since the 
generic elliptic fiber admits one-cycles which pinch on the
location of the seven-branes. This construction is natural for 
$\cS$, since one has more than one 
brane in the presence of a non-Abelian gauge group. 
These branes have been moved apart once one blows up 
the singularity introducing new divisors $\hat D_i$ and corresponding 
two-forms $\tw_i$. 
Using the complex structure of $\hat X_4$ to split $H^{1,0}(\cS) \oplus H^{0,1}(\cS)$
this yields a map $H^{1,0}(\cS) \hookrightarrow H^{2,1}(\hat X_4)$. 
Locally,  one can write these $(2,1)$ forms as
\beq \label{local_xii}
  \Psi_{a\, i} \cong  \gamma_a \wedge \tw_i, 
\eeq
where $\tw_i$ are $(1,1)$ forms on $\hat X_4$ already introduced above. 
The local expressions for $ \Psi_{a\,i}$ might patch together 
to harmonic forms representing elements of $H^{2,1}(\hat X_4)$. 
The Wilson line scalars $\tN^{a\,i}$ in the M-theory reduction carry 
an extra index labeling the seven-brane.
Recall that in section \ref{F-dual_KK} we discussed the $h^{2,1}(B_3)$
vectors which correspond to scalars $\cN_\kappa$ in three dimensions.
Here we will set $h^{2,1}(B_3)=0$ and show that the remaining fields 
in $H^{2,1}(\hat X_4)$ correspond to complex scalars $\tN^a$ in four dimensions.
We will discuss the general case in appendix~\ref{allH3}, where we include the full split 
\beq  \label{NIsplit}
  N^I = (\cN_\kappa,\tN^a)\ ,
\eeq
with $\cN_\kappa$ descending to vectors and $\tN^a$ descending to scalars in the F-theory lift.

There is also a second set of scalars which is counted by elements in 
$H^{2,1}(\hat X_4)$. Namely, also degrees of freedom from the 
R-R and NS-NS two-form $C_2,B_2$, combined into a complex $G_2$ in \eqref{monodr-fields}, 
can patch together and yields scalar fields in the dimensional reduction.
To make this more precise, let us introduce the $\nu^{(\kappa)}$ complex one-forms on the elliptic fiber
\beq
   \nu^{(\kappa)} = \tfrac{i}{2} (\I \tau)^{-1} (dx + \bar \tau dy)\ .
\eeq
These are defined in a patch $\cU_{\kappa}$ on $B_3$ labeled by $\kappa$. 
On the overlap $\cU_{\kappa} \cap \cU_{\lambda}$ of two such patches 
these one-forms can transform with an $Sl(2,\bbZ)$ transformation 
as evaluated using \eqref{tau_trans}.
We denote by $C^{(\kappa)}_2,B_2^{(\kappa)}$ the two-forms in the 
patch $\cU_\kappa$.   
Locally one can now introduce a three-form 
\beq \label{BC_patch}
   B^{(\kappa)}_2 \wedge dx + C^{(\kappa)}_2 \wedge dy = G^{(\kappa)}_2 \wedge \nu^{(\kappa)} + \bar G^{(\kappa)}_2 \wedge \bar \nu^{(\kappa)}\ . 
\eeq
This three-form transforms invariantly when moving from patch to patch on the base $B_3$, and 
hence can lead to zero modes which have to be included in an F-theory reduction. 

Clearly, the expression \eqref{BC_patch} will be part of the M-theory three-form 
when describing F-theory via the M-theory lift. Hence, we have to perform a reduction 
very similar to the one of section \ref{F-dual_KK}. 
We thus expand $C_3$ as
\beq \label{C3expansion_2}
  C_3 = A^\cA\wedge e_\cA + a^a\, \alpha_a + b_a \, \beta^a 
      = A^\cA \wedge e_\cA + \bar \tN^a\, \Psi_a + \tN^a\, \bar \Psi_{a}\ ,   
\eeq
where $(a^a,b_a)$ are three-dimensional real scalars which combine into complex scalars $\tN^a$. 
The three-forms
$(\alpha_a,\beta^a),\, a=1,\ldots,h^{2,1}(\hat X_4)$ in \eqref{C3expansion_2} comprise a 
basis of $H^{3}(\hat X_4,\bbZ)$. In contrast to the basis $(\alpha_\kappa,\beta^\kappa)$ introduced in 
section \ref{F-dual_KK}, the basis $(\alpha_a,\beta^a)$ is not canonically symplectic, since it 
is generically not supported only on one divisor in $\hat X_4$. However, if $\tN^a$ correspond 
exclusively to Wilson line degrees of freedom on $\cS$, one can make contact with the
discussion from the begin of this subsection.
The Wilson lines $\tN^{a\, i}$ are labeled with an 
extra index $i$ counting the number of branes on $\cS$. The corresponding 
$(2,1)$ forms $\Psi_{a\,i}$ are near $\cS$ of the form \eqref{local_xii}.
However, in the following we will keep the analysis general and only comment on the 
Wilson line case at the end of the section.

\begin{table}[h!] 
\begin{center}
\begin{tabular}{|c||c|c|} \hline  
 \rule[-0.3cm]{0cm}{0.9cm}
       3-dim~multiplet  & \multicolumn{2}{c|}{4-dim~F-theory}    \\ \hline
 \rule[-0.3cm]{0cm}{0.9cm}
 \multirow{2}{2cm}{$\tN^a$}  & \multirow{2}{7cm}{\\[-.2cm] \quad $h^{2,1}(\hat X_4)-h^{2,1}(B_3)$ chiral multiplet \quad }& \quad Wilson line scalars \hspace*{.2cm} \\ \cline{3-3}
 \rule[-0.2cm]{0cm}{0.8cm}
 &  & $B_2,C_2$ scalars \\
 \hline  
\end{tabular} 
\caption{\textit{Chiral multiplets arising in the 
F-theory lift of complex scalars in $C_3$.}} \label{F-theory_lift2}
\end{center}
\end{table}

As in section \ref{F-dual_KK} the complex structure defining $\tN^a$ 
is induced by the complex structure on $\hat X_4$, as captured 
by the $(2,1)$-forms $\Psi_a$. However, one now choses a basis to 
expand $\Psi_a$ as 
\beq \label{Psi_exp}
  \Psi_a =  \alpha_a  - i   \bar f_{ab} \beta^b \ , \qquad \I \Psi_a =-  \R f_{ab} \beta^b\ ,
\eeq
for a holomorphic function $f_{ab}$ of the complex structure moduli $z^\cK$.
These are the analogs of the forms introduced in \eqref{gamma_exp}.
As an immediate consequence of \eqref{C3expansion_2} and \eqref{Psi_exp} one concludes that 
\beq
  \tN^a =  \tfrac{1}{2}\R f^{ab}(\bar f_{ab}\, a^b + i b_a)\ ,
\eeq
with $\R f^{ab}\equiv (\R f_{ab})^{-1}$ being the inverse of the real part of $f_{ab}$. 
The couplings of the fields $\tN^a$ are, as in \eqref{d_AIJ}, captured by the complex structure dependent function 
\beq \label{def-dcAab}
   d_{\cA a\bar b} = i\int_{\hat X_4} \omega_\cA \wedge \Psi_a \wedge \bar \Psi_b\ .
\eeq
Note that in contrast to the case of section \ref{F-dual_KK} the $\Psi_a$ do not have all indices on $B_3$ 
since we have assumed $h^{2,1}(B_3)=0$. Hence, the 
only non-vanishing couplings are actually $d_{\alpha a \bar b}$, since only $\omega_\alpha$
has only indices on the base $B_3$. 
Hence, a computation similar to the one leading to \eqref{correctedtildeK_vec}, yields the 
kinetic potential  
\beq \label{correctedtildeK}
  \tilde K^{\rm M} = \tilde K_o^{\rm M}(R,L,\xi) + L^\alpha d_{\alpha a \bar b}(z,\bar z)\,\R \tN^a \R \tN^b + \ldots\ ,
\eeq
where $\tilde K_o^{\rm M}(R,L,\xi)$ is the original kinetic potential independent of the 
fields $\tN^a$, and $L^\alpha$ are the scalars in the three-dimensional vector multiplets $(A^\alpha,L^\alpha)$
which dualize to four-dimensional complex scalars $T_\alpha$. 

In considering the F-theory lift to four dimensions one realizes 
that \eqref{correctedtildeK} cannot be the complete correction. To see this, one notes 
that the real part of the holomorphic gauge coupling function $f_{ij}$ has to obey \eqref{holom_fij_MF}.
Since the corrections term in \eqref{correctedtildeK} modifies the definition of the 
coordinates $T_\alpha$, holomorphy of the four-dimensional gauge coupling implies that 
that \eqref{correctedtildeK} is missing a correction proportional to $\xi^2$. To ensure \eqref{diff_eq}
with $\tilde f=0$,  one finds that the correct expression for $\tilde K^{\rm M}$ is
\beq
   \tilde K^{\rm M} = \tilde K^{\rm M}_o(R,L,\xi) + (L^\alpha - R^{-1}\, C^\alpha_{ij}\, \xi^i \xi^j )\, d_{\alpha a \bar b}(z,\bar z)\,\R \tN^a \R \tN^b\ .
\eeq
Note that the new term in $\tilde K^{\rm M}$ which is linear in $L^\alpha$ is removed by the Legendre 
transform to coordinates $T_\alpha$ and kinetic potential $\mathbf{K}^{\rm M}$ as in 
\eqref{Talpha_Leg1} and \eqref{KM_claim2},\eqref{holom_fij_MF}. More precisely, one has
\beq \label{Talpha_Leg2}
  T_\alpha =  \partial_{L^\alpha}\tilde K^{\rm M}_o(R,L,\xi) + d_{\alpha a\bar b}(z,\bar z)\,\R \tN^a \R \tN^b + i\rho_\alpha\ ,
\eeq 
and 
\beq \label{KMwithWilson}
   \mathbf{K}^{\rm M} = \log R + K^{\rm F} - \frac{1}{2R} \R T_{S'} \, \xi^i \xi^j + \cO(R)\ .
\eeq
In this expression $K^{\rm F}$ is readily evaluated from the original 
$K^{\rm F}_o$ by replacing $T_\alpha \rightarrow T_{\alpha} +d_{\alpha a\bar b} \R \tN^a \R \tN^b$.
It is straightforward to evaluate this expression for the $\tilde K_o^{\rm M}$ given in \eqref{tildeKM1}.

In the F-theory lift of the expressions \eqref{KMwithWilson} one uses again the
fact that gauge fields descent to a non-Abelian theory removing the indices $i,j$ and
replacing the gauge coupling function as in \eqref{TS'}. Moreover, 
if all $\tN^a$ correspond to Wilson line moduli $\tN^{a\,i}$ the expression
\eqref{Talpha_Leg2} is lifted to 
\beq \label{RTwithWilson}
  \R T_\alpha =  \partial_{L^\alpha}\tilde K^{\rm M}_o(R,L,\xi) + d_{\alpha a \bar b} \text{Tr}(\R \tN^a\, \R \tN^b) \ ,
\eeq
where the four-dimensional $\tN^a$ transform in the adjoint of $G$, and we have inserted 
\beq 
   d_{\alpha a \bar b} = i \int_{\cS} \omega_\alpha \wedge \gamma_a \wedge \bar \gamma_b\ ,
\eeq
as can be evaluated using \eqref{local_xii}, \eqref{def-dcAab}, 
and an identity similar to \eqref{tw_identity}. It is not hard to check that this lift leads to 
the correct local expression for the kinetic terms of the Wilson line moduli 
studied in \cite{Beasley:2008dc}. Furthermore, \eqref{KMwithWilson} together with \eqref{RTwithWilson} 
yields the correct expression in the orientifold limes \cite{Jockers:2004yj} and 
orbifold limits \cite{Lust:2004cx,Blumenhagen:2006ci}.

\section{Conclusions}

In this paper we studied the four-dimensional effective action of 
F-theory compactified on an elliptically fibered Calabi-Yau fourfold  $X_4$. Many of the 
important equations and results of this paper are summarized in appendix \ref{summary_app}.
The main 
tool was to use a non-trivial scaling limit to lift the three-dimensional supergravity theory obtained by compactification
of M-theory on $X_4$. We have shown explicitly how 
the massless Kaluza-Klein modes arising in the reduction of the M-theory three-form and the K\"ahler 
form along the elliptic fiber of $X_4$ encode the massless metric degrees of freedom along an $S^1$ 
used in a compactification from four to three dimensions. The guiding principle to formulate 
the M- to F-theory lift has been to demand finiteness of M5-brane actions wrapped on vertical divisors.
An M5-brane wrapped on the base $B_3$ has infinite action in the F-theory limit as it scales as the 
square of the $S^1$ radius.

The M- to F-theory lift can be extended 
to singular elliptic fibrations if the singularity has been resolved by blow-up yielding a fourfold $\hat X_4$. 
The degrees of freedom induced by the new blow-up forms correspond to the massless degrees of freedom in 
the reduction of four-dimensional $U(1)$ vectors. The gauge-group enhances to a non-Abelian 
group due to massless M2-branes on resolving cycles of vanishing volumes. 
In the four-dimensional F-theory picture this
corresponds to a stack of seven-branes with non-Abelian gauge group on their world-volume. 
We considered the F-theory limit using the large volume expressions for the resolved 
Calabi-Yau fourfold $\hat X_4$ to compute the leading gauge-coupling function 
for the seven-branes. It is of interest to study the corrections to these expressions. 
In particular, the splitting of the gauge-couplings in the orientifold limit observed in ref.~\cite{Blumenhagen:2008aw},
has to be investigated in the M- to F-theory lift. However, it is crucial to note that 
these corrections depend on the dilaton-axion $\tau$, and hence are expected to be generalized 
via a holomorphic function of the complex structure moduli as in section~\ref{matter_couplings}.

{}From the M-theory reduction we were able to extract the four-dimensional K\"ahler potential 
and gauge coupling functions. In three dimensions both are encoded by a single function, the kinetic
potential $\mathbf{K}^{\rm M}$, which has to be 
expanded around the F-theory limit in 
the  K\"ahler moduli space. The expressions found in the F-theory reduction specialize to the
results found in the orientifold limit in refs.~\cite{Grimm:2004uq,Jockers:2004yj}. It will 
be crucial to examine corrections to these leading order results. While the KKLT scenario \cite{Kachru:2003aw} 
of stabilizing K\"ahler moduli has an implementation in F-theory, it remains to be shown how the large volume 
compactifications of \cite{Balasubramanian:2005zx} can be realized. This is again due to the fact that the corrections 
used in \cite{Balasubramanian:2005zx} dependent on the dilaton-axion and lift non-trivially to F-theory. 

In addition to the seven-brane gauge theories we have also investigated the 
gauge dynamics of the Abelian gauge theory arising in the reduction of the 
R-R four-from. Massless vectors arise in the F-theory reduction if the 
base of the fourfold $\hat X_4$ admits harmonic three-forms. The 
gauge-coupling function is determined by a holomorphic function 
$f_{\kappa \lambda}(z)$ encoding the fibration of a torus bundle $\mathbb{T}$
over the complex structure moduli space of $\hat X_4$. It is an interesting 
task to compute this holomorphic function explicitly for  Calabi-Yau fourfold examples.
There will be special loci in the complex structure moduli space at which this Abelian 
theory enhances to a non-Abelian gauge group, just as in Seiberg-Witten theory.
To study the R-R gauge theory at various points in open-closed complex structure 
moduli space might yield new insights for $\cN=1$ gauge theories. In particular,
it will be interesting to explore how the singularities for non-Abelian seven-branes
are seen by the R-R gauge theory.

In the last section we have focused on matter corresponding to seven-brane deformations 
and Wilson line moduli which transform in the adjoint of the non-Abelian gauge group. 
Again we were able to determine their couplings in the $\cN=1$ K\"ahler potential. This allowed 
us to derive the matter D-term including the flux correction. The Wilson line moduli 
non-trivially correct the K\"ahler moduli. 
We argued that in M-theory these fields arise as zero modes 
along harmonic three-forms with one leg on the seven-brane surface $\cS$. 
It will be a crucial task to extend the analysis to localized charged matter arising 
at intersections of the seven-brane on $\cS$ with other seven-branes. These fields 
can be chiral if fluxes on the world-volume are turned on. It is conceivable that the
results of section \ref{matter_couplings} naturally extend to this case if one considers fields 
$\varphi',A'$ transforming in the adjoint of $G^{\rm max}$ introduced in this section. However,
the zero mode conditions appear to mix contributions from the two sectors, and it will be 
interesting to establish a formulation where this mixing is canonically captured.
Let us conclude by noting that questions concerning moduli stabilization and 
chirality both require a detailed understanding of fluxes in F-theory, 
and the full picture needs yet to be explored.\footnote{For a recent discussion using heterotic/F-theory duality, see refs.~\cite{Grimm:2009ef,Grimm:2009sy,Jockers:2009ti,Marsano:2010ix}.}

\section*{Acknowledgments}

I am grateful to R.~Blumenhagen, M.~Cheng, F.~Denef, O.~Hohm, D.~Klevers, A.~Klemm, H.~P.~Nilles, T.~Weigand, and  T.~Wotschke for enlightening discussions
and useful comments on the draft. 
I like to thank the Kavli Institute for Theoretical Physics
and the Max Planck Institute for Physics in Munich for hospitality during parts of this work. 
This research was supported in part by the SFB-Transregio 33 ``The Dark Universe'' by the DFG.

\vspace*{1cm}

\appendix

\noindent {\bf \LARGE Appendices}

\section{Summary of results} \label{summary_app}

In this appendix we summarize some of the key results of this work. 
Most of the details are skipped and can be found in the main text.
The general results for the Kaluza-Klein reduction from four to three dimensions 
can be found in appendix \ref{scalar_vec_duality}.

\subsubsection*{General expressions for the M-theory reduction}

Considering M-theory on a resolved Calabi-Yau fourfold $\hat X_4$ the
K\"ahler potential in three dimensions is
\bea \label{general_KMsum}
   K^{\rm M}(z,T,N) &=& K^{\rm cs}(z) - 3\log \cV \\
   &=& 
   - \log \int_{\hat X_4} \Omega \wedge \bar \Omega - 3 \log \int_{\hat X'_4} \Omega' \wedge \bar \Omega'\ ,\nn 
\eea
as discussed in \eqref{class_KM}, \eqref{KX'} and \eqref{def-Kcs_gen}. The quantum volume $\cV$
is determined by mirror symmetry, where $\hat X'_4$  is the
mirror of $\hat X_4$, and $\Omega'$ is the mirror $(4,0)$ form. $K^{\rm M}$
depends on the complex structure moduli $z^\cK$ through $\Omega$.
The second term has to be evaluated on a real Lagrangian slice of dimension 
$h^{1,1}(\hat X_4) = h^{3,1}(\hat X'_4)$. We choose real 
coordinates $v^\cA$ to parameterize this slice.
The real part of the K\"ahler moduli $\R \, T_\cA$ are defined as real parts
of the mirror periods $\Pi^{(3)}_\cA(v)$ corresponding
to the mirror of six-cycles in $\hat X_4$ (see e.g.~\eqref{red-T0}, \eqref{Talpha_Leg2}):
\beq \label{general_Tsum}
   T_\cA = \R\, \Pi^{(3)}_\cA(v) + d_{\cA I J}(v,z) N^I \R N^J+ i \tilde \rho_\cA\ , 
\eeq
where $d_{\cA I J}$ is in general a function of $v^\cA,z^\cK$. Chiral multiplets $N^I$ 
with axion-like $\I N^I$ appear quadratically in $T_\cA$. There can be other moduli 
corresponding to further matter multiplets. The K\"ahler potential 
\eqref{general_KMsum} has to be evaluated as a function of $z^\cK,T_{\cA},N^I$.
In particular, one has to solve
\beq
  2\, \R\, \Pi^{(3)}_\cA(v)  = T_\cA + \bar T_\cA -   2\, d_{\cA I J}(v,z) \R N^I \R N^J
\eeq
for $v^\cA$ and insert the result into \eqref{general_KMsum}.
If $N^I$ transforms as an adjoint under a non-Abelian gauge group one 
has to replace $ \R N^I \R N^J \rightarrow \text{Tr}(\R N^I \R N^J)$.
Depending on the type of six-cycle 
one has the split
\beq \label{TcA_split}
   T_\cA = (T_0, T_\alpha , T_i )\ ,
\eeq
where $T_0$ corresponds to the base $B_3$, the $T_\alpha$ correspond to vertical divisors $D_\alpha$, and 
the $T_i$ are associated with resolution divisors $\hat D_i$ (subsection \ref{Ell+seven}).
If $\I T_\cA$ has a shift symmetry all $T_\cA$ can be dualized to three-dimensional vector multiplets
\beq
  (L^\cA ,A^\cA)\ =\ \big((R,A^0)\, , \ (L^\alpha, A^\alpha)\, , \ (\xi^i,A^i) \big)\ ,  
\eeq
in accord with the split \eqref{TcA_split}. This dualization corresponds to performing a 
Legendre transform 
\beq
   \tilde K^{\rm M}(L,z,M) = K^{\rm M} - \frac{1}{2} (T_\cA + \bar T_\cA) L^\cA \ , \qquad  \frac{\partial K^{\rm M}}{\partial T_\cA}= -\frac{1}{2} L^\cA\ .
\eeq
$\tilde K^{\rm M}$ is a kinetic potential encoding the kinetic terms of the complex 
scalars $z^\cK,N^I$ as well as the vector multiplets $(L^\cA,A^\cA)$ in the 
three-dimensional action \eqref{kinetic_lin_gen_1}.
It is often useful to work with $\tilde K^{\rm M}$ since the dependence on $L^\cA$ 
can be much simpler as the $T_\cA$-dependence of $K^{\rm M}$. 
This is the case at large volume.

\subsubsection*{M-theory reduction at large volume}

If all volumes in $\hat X_4$ are large the classical expressions for the 
mirror periods are inserted into \eqref{general_KMsum} and \eqref{general_Tsum}: 
\beq \label{largeVsum}
   \cV = \frac{1}{4!}\int J^4\ , \quad \qquad T_\cA = \frac{1}{3!} \int_{D_\cA} J^3  + d_{\cA I J}(z) N^I \R N^J+ i \tilde \rho_\cA\ .
\eeq 
For an elliptic fibration with a resolved $G$-singularity over $\cS$ 
one has the following conditions for the quadruple intersections:
\bea
 && D_\alpha \cdot D_\beta \cdot D_\gamma \cdot D_\delta = \hat D_i \cdot D_\beta \cdot D_\gamma \cdot D_\delta = 0 \ ,\\
 && (\hat D_{i} \cdot \hat D_{j} + C_{ij} \, S' \cdot B_3) \cdot D_\alpha \cdot D_\beta = 0 \ ,
\eea
where $C_{ij}$ is the Cartan matrix for $G$.
The $d_{\cA IJ}$ in \eqref{largeVsum} are independent of $v^\cA$ at large volume, and given by 
\beq
   d_{\cA IJ} = i \int_{\hat X_4} \omega_\cA \wedge \psi_{I}\wedge \bar \psi_{J}\ ,
\eeq
where $\psi_{I}$ are $(2,1)$ forms on $\hat X_4$. These $(2,1)$-forms vary over 
the complex structure moduli space. The complex structure dependence 
is encoded by a holomorphic function $f_{IJ}(z)$ as in \eqref{psi_exp} and \eqref{Psi_exp}.
The kinetic potential $\tilde K^{\rm M}$ at large volume is (see \eqref{tildeKM1} and \eqref{correctedtildeK})
\bea \label{tKM_simplesum}
   \tilde K^{\rm M} &=& \log[\tfrac{1}{6} R L^\alpha L^\beta L^\gamma \cK_{\alpha \beta \gamma} - \tfrac{1}{4} \xi^i \xi^j \, C_{ij} \, L^\alpha L^\beta \cK_{S|\alpha \beta}+\cO(R^3,\xi^3)] + K^{\rm cs}(z) \nn \\ 
   && + L^\cA d_{\cA IJ} \R N^I \R N^J + \cO(\xi^2,R^2) \ .
\eea
Using the index structure of the $(2,1)$-forms on $\hat X_4$ 
one splits $N^I = (\tN^a,\cN_\kappa)$ as in \eqref{NIsplit}, and appendix \ref{allH3}, and has 
\beq
    L^\cA d_{\cA IJ} N^I N^J  = L^\alpha d_{\alpha ab}  \tN^a  \tN^b + R d_{0}^{\ \kappa \lambda} \cN_\kappa \cN_\lambda + 2 \xi^i d_{i a}^{\ \ \kappa} \tN^a \cN_\kappa\ .
\eeq
The $\psi^\kappa$ in $d_{0}^{\ \kappa \lambda}$ have all indices in the base $B_3$ as in \eqref{psi_exp}.
The $\Psi_a$ are the remaining $(2,1)$ forms in \eqref{Psi_exp}.

\subsubsection*{F-theory lift, moduli matching, and the $\cN=1$ characteristic data} 

The F-theory limit is a limit in K\"ahler moduli space given by 
\bea \label{F-theorylimit_sum}
  R \ \rightarrow \ 0\ , \qquad L^\alpha \ \rightarrow L^\alpha_{\rm b} \ , \qquad \xi^i/R \ \rightarrow \ 0\ ,\\
  \R T_0 \rightarrow \ \infty\ , \qquad T_\alpha \ \rightarrow T_\alpha^{\rm b} \ , \qquad \R T_i \ \rightarrow \ 0\ .\nn
\eea
This limit sets the background values of the fields $R,\xi^i$. Their variations appear in the 
effective action. 
The three-dimensional M-theory compactification expanded around this limit has to be compared 
with the dimensional reduction of a general four-dimensional $\cN=1$ supergravity theory
reduced to three dimensions on an $S^1$ of radius $r$.  
The elliptic fiber volume $R$ is identified as $R = r^{-2}$.  
The three-dimensional fields lift non-trivially to four dimensions as summarized in 
table \ref{F-theory_spec_tab}. 
\begin{table}[h!] 
\begin{center}
\begin{tabular}{|c||c|c|} \hline  
 \rule[-0.3cm]{0cm}{0.9cm}
       3-dim.~multiplet  & \multicolumn{2}{c|}{4-dim.~F-theory}    \\ \hline
 \rule[-0.3cm]{0cm}{0.9cm}
 \multirow{2}{2cm}{$(L^{\cA},A^\cA)$}  & \quad $h^{1,1}(\hat X_4)-h^{1,1}(B_3)-1$ vector multiplets \quad & \quad $(\xi^i,A^i)\rightarrow A^i$ \hspace*{.2cm} \\ \cline{2-3}
 \rule[-0.2cm]{0cm}{0.8cm}
 &  $h^{1,1}(B_3)$ chiral multiplets & $(L^\alpha,A^\alpha)\rightarrow T_\alpha$ \\ \cline{2-3}
 \rule[-0.2cm]{0cm}{0.8cm}
 &  extra dimension & $(R,A^0)\rightarrow (g_{33},g_{\mu 3})$ \\ 
 \hline  
 \hline
 \rule[-0.3cm]{0cm}{0.9cm}
 $N^I$ 
 & \quad $h^{2,1}(B_3)$ vector multiplets \quad & \quad $\Nk\rightarrow V^\kappa$ \hspace*{.2cm} \\ \cline{2-3}
 \rule[-0.2cm]{0cm}{0.8cm}
 &  $h^{2,1}(\hat X_4)-h^{2,1}(B_3)$ chiral multiplets & $\Na \rightarrow \Na$ \\ 
 \hline  
 \hline
  \rule[-0.2cm]{0cm}{0.8cm} $z^\cK$  & \quad $h^{3,1}(\hat X_4)$ chiral multiplets \quad & \quad $z^\cK\rightarrow z^\cK$ \hspace*{.2cm}   \\ 

 \hline
\end{tabular} 
\caption{\textit{The four-dimensional F-theory spectrum without matter fields.}} \label{F-theory_spec_tab}
\end{center}
\end{table} 

Since among the $T_\cA$ in \eqref{TcA_split} only $T_\alpha$ lifts to a complex scalar in four dimensions, the key object to 
work with is the kinetic potential 
\beq \label{bfK_expsum}
  \mathbf{K}^{\rm M}(T_\alpha,z,N|\xi,R) = \tilde K^{\rm M} - \tfrac12 (T_\alpha + \bar T_\alpha) L^\alpha\ , \qquad \R\, T_\alpha = \partial_{L^\alpha} \tilde K^{\rm M}\ .
\eeq
This potential depends on the scalars $\xi^i,R$ which are in vector multiplets. The $\cN_\kappa$ lift to four-dimensional vectors, 
but appear as complex scalars in $\mathbf{K}^{\rm M}$. One expands $\mathbf{K}^{\rm M}$ around the  F-theory limit \eqref{F-theorylimit_sum}, $\R \cN_\kappa \rightarrow 0$,
as 
\beq
     \mathbf{K}^{\rm M} = \log R + K^{\rm F} (z,T,\tN)|_* + \tfrac{1}{2} \tilde K^{\rm M}_{\xi^i \xi^j}|_{*} \xi^i \xi^j 
                                                          + \tfrac{1}{2} \tilde K^{\rm M}_{\R \cN_\kappa \R \cN_\lambda}|_{*} \R \cN_{\kappa} \R \cN_{\lambda} + \ldots \ ,
\eeq
where $|_*$ indicates evaluation in the strict limit. There is no linear term in $\cN_\kappa$ due to its quadratic appearance in \eqref{general_Tsum}.
Linear terms in $\xi^i$ have to be absent to ensure match with a kinetic potential obtained by dimensional reduction.
Using appendix \ref{scalar_vec_duality} a \textit{general} $\mathbf{K}^{\rm M}$ obtained by reduction is 
\beq \label{bfK_redsum}
 \mathbf{K}^{\rm M} = \log R + K(M,\bar M) - \tfrac12 R^{-1} \Delta_{ij}\, \xi^i \xi^j  
                      + \tfrac12 R\, \Delta^{\kappa \lambda} \, \R \cN_\kappa \R \cN_\lambda 
                      -\, \Delta^\lambda_i \xi^i \, \R \cN_\lambda  \ ,
\eeq
where $K$ is the four-dimensional K\"ahler potential and  $\Delta$ is given in \eqref{def-deltaapp}.
For $\Delta^\lambda_i =0$ one has: $ \Delta_{ij} = \R f_{ij}$, $\Delta^{\kappa \lambda} = \R f^{\kappa \lambda}$, 
where $f_{ij}$ and $f_{\kappa \lambda}$ are holomorphic gauge coupling functions.
In order that $T_\alpha$ can arise holomorphically in $f_{ij}$ the kinetic 
potential $\tilde K^{\rm M}$ has to satisfy 
\beq \label{diff_eqsum}
    \big(\cC^\alpha_{ij} \partial_{L^\alpha}  +  R \partial_{\xi^i} \partial_{\xi^j} \big)\tilde K^{\rm M} 
    =  \R\, \tilde f_{ij}(M)  \ , \qquad \text{for} \ R,\xi^i/R \rightarrow 0\ .  
\eeq
Here $\tilde f_{ij}$ is a holomorphic function which can in general appear in the gauge-coupling 
function. It was found to be zero at leading order. 

Since the indices $i,j$ parameterize 
the blow-up divisors of a $G$ singularity over $\cS$, they label Cartan $U(1)'s$ in $G$. 
In the F-theory limit \eqref{F-theorylimit_sum} the indices $i$ run over all generators, 
and the gauge group enhances as $U(1)^{rk(G)} \rightarrow G$, due to light M2-branes.
Similarly, one can treat matter deformations $\varphi'$ and Wilson line moduli transforming in the 
adjoint of $G$, as discussed in section \ref{matter_couplings}. To display the 
formulas, we assume that all $\tN^a$ are Wilson line moduli, since otherwise we have to introduce 
new indices and split the set $\tN^a$. 
For the simple $\tilde K^{\rm M}$ given in \eqref{tKM_simplesum} one compares \eqref{bfK_expsum} with \eqref{bfK_redsum}. 
One then finds  
\bea
   K^{\rm F}(z,T,\varphi',\tN) &=&  - 2 \log \cV_{\rm b} - \log \big[\int_{\hat X_4} \Omega \wedge \bar \Omega - \int_\cS \text{Tr}(\varphi' \wedge \bar \varphi') \big]\ ,\\\
                     T_\alpha &=& \frac{1}{2!} \int_{D_\alpha^{\rm b}} J_{\rm b} \wedge J_{\rm b} + d_{\alpha a b}(z) \text{Tr} ( \tN^a \R \tN^b) + i \tilde \rho_\alpha\ ,
\eea
together with the gauge coupling functions 
\beq
  f_{G} = T_{S'} \ , \qquad \qquad f^{\rm RR}_{\kappa \lambda} = f_{\kappa \lambda}(z)\ . 
\eeq
Note that the couplings $d_{\alpha a b}(z)$, $ f_{\kappa \lambda}(z)$ as well as the correction $\varphi' \wedge \bar \varphi'$ depend on 
all complex structure moduli of $\hat X_4$. In particular, this includes couplings to seven-brane moduli.
Note that various additional corrections can be computed using this formalism. For example, we have indicated volume 
corrections to $K^{\rm F}$ in \eqref{Delta_cV}. Finally, the superpotential was given in \eqref{full_WF}, \eqref{Yukawas} and 
reads 
\beq
    W^{\rm F} = \int_{\hat X_4} G_4 \wedge \Omega(z) + \int_{\cS} \text{Tr}(F_{\rm flux} \wedge \varphi') + Y_{ab\nu} \text{Tr}(\tN^a \tN^b \varphi'^\nu) \ , 
\eeq
while the leading order D-term is computed in section \ref{Adjoint_deforms} and reads
\beq
  D_{G} = \frac{1}{4 \cV_{\rm b}} \int_{\cS} J_{\rm b} \wedge F_{\rm flux} + \frac{i}{\int_{X_4} \Omega \wedge \bar \Omega}\int_{\cS}  [\varphi',\bar \varphi']\ .
\eeq

\section{$4d \rightarrow 3d$ reduction and scalar-vector duality }\label{scalar_vec_duality}

In this appendix we discuss the reduction 
of the bosonic $\cN=1$ action \eqref{N=1action} to 
three space-time dimensions on a circle $S^1$ of radius $r$ (see, e.g., refs.~\cite{Haack:1999zv}). 
The four-dimensional metric, and the four-dimensional vectors split 
as 
\beq \label{3d-Ansatz}
  g^{(4)}_{\mu \nu} = \left(
  \begin{array}{cc}
   g^{(3)}_{pq} + r^2 A_p^0 A_q^0& r^2 A_q^0\\ r^2 A_p^0 & r^2
  \end{array} 
  \right)\ , \qquad A^\Lambda_\mu = (A^\Lambda_p+A^0_p\, \zeta^\Lambda, \zeta^\Lambda)\ ,
\eeq 
where $A^0_p,\ A^\Lambda_p,\ p=0,1,2$ are vectors and $\zeta^\Lambda$ as well as $r$ are
scalars in three dimensions. Note that we will use the same symbol $A^\Lambda$ for 
four- and three-dimensional vectors. However, it should be clear by the context whether 
we are working in four or three dimensions.  
Performing the Kaluza-Klein reduction of the action \eqref{N=1action} and 
employing a Weyl rescaling to the three-dimensional Einstein frame, the 
three-dimensional action can be brought into the standard form 
\bea\label{kinetic_lin_gen}
  S^{(3)} &=& \int-\tfrac{1}{2}R_3 *\mathbf{1} - 
  \tilde K_{I \bar {J}}\, dM^I \wedge * d \bar M^{J}
  + \tfrac{1}{4} \tilde K_{\hat \Lambda \hat\Sigma}\, 
  d\xi^{\hat \Lambda}\wedge * d\xi^{\hat \Sigma} \nn\\ 
  && - \tfrac{1}{4} \tilde K_{\hat \Lambda \hat\Sigma}\, F^{\hat\Lambda} \wedge * F^{\hat\Sigma}
     + \,  F^{\hat \Lambda} \wedge \I (\tilde K_{\hat\Lambda I}\, dM^I)\ ,
\eea
where the kinetic terms of the vectors and scalars are determined by 
a single real function, the kinetic potential $\tilde K(M^I,\bar M^J|\xi^{\hat \Lambda})$, as
$\tilde K_{I {\bar J}} = \partial_{M^I} \partial_{\bar M^J} \tilde K$, $\tilde K_{\hat \Lambda \hat \Sigma} = \partial_{\xi^{\hat \Lambda}} \partial_{\xi^{\hat \Sigma}} \tilde K$, and 
$\tilde K_{\hat \Lambda I} = \partial_{\xi^{\hat \Lambda}} \partial_{M^I} \tilde K$.
In order to do that, we identify
\beq \label{identify}
  R = r^{-2}\ ,\qquad \xi^{\hat \Lambda} = (R,R \zeta^\Lambda)\ ,\quad A^{\hat \Lambda} = (A^0,A^\Lambda)\ .
\eeq
It is straightforward to determine the kinetic potential $\tilde K$. Clearly, it will contain 
the four-dimensional K\"ahler potential $K(M^I,\bar M^J)$, and additional terms 
encoding the kinetic terms of the vector multiplets $(A^{\hat \Lambda},\xi^{\hat \Lambda})$. 
Explicitly it takes the form\footnote{Note that we have included a factor $1/2$, by rescaling the vector fields.} 
\beq \label{d=3_red_potential}
  \tilde K = K(M,\bar M) + \log R - \frac{1}{2 R} \R f_{\Lambda \Sigma}(M)\, \xi^{\Lambda} \xi^{\Sigma}\ . 
\eeq
It is important to stress that in three dimensions massless vectors are 
dual to real scalars with Peccei Quinn shift symmetries. However, in the 
kinetic potential \eqref{d=3_red_potential} one can still distinguish the four-dimensional 
origin of the term by considering the power $n$ of the $R^n$ pre-factor.
In fact, four-dimensional scalars carry no pre-factor in the $D=3$ kinetic potential 
and action, while three-dimensional scalars $\zeta^\Lambda \equiv \xi^\Lambda/R$ 
which arise from $D=4$ vector multiplets carry a pre-factor $R^{-1}$ in \eqref{d=3_red_potential}. 

To study the F-theory reduction it turns out to be convenient to dualize some
of the vectors $A^\Lambda$ in \eqref{identify} into scalars. Note that only 
those vectors are dualizable which do not gauge 
any scalar fields in the effective theory. For example, in an F-theory reduction these 
correspond to the $U(1)$ vectors $A^\kappa$ which arise in the reduction of the
R-R four-form $C_4$ as discussed in section \ref{RRsection}. We thus split 
\beq
  A^\Lambda = (A^\kappa,A^i)\ , \qquad \xi^\Lambda = (\xi^\kappa,\xi^i)\ , \qquad f_{\Lambda \Sigma} = (f_{ij},f_{\kappa j},f_{\kappa \lambda})\ .
\eeq
To dualize the vectors $A^\kappa$ into scalars $\tilde \xi_\kappa$ one adds a Lagrange 
multiplier term $\propto F^\kappa \wedge d \tilde \xi_\kappa$ to the effective 
action \eqref{kinetic_lin_gen} and eliminates 
$F^\kappa$ by its equation of motion. At the level of the kinetic potential $\tilde K$
and coordinates this amounts to performing a Legendre transformation with respect to $\tilde K$. 
The scalars $(\xi^\kappa,\tilde \xi_\kappa)$ then combine into a complex coordinates $\cN_\kappa$.
More precisely, one introduces new complex coordinates
\beq \label{N=dtildeK}
   \cN_\kappa = - \partial_{\xi^\kappa} \tilde K - i\tilde \xi_\kappa  \ ,
\eeq
and transforms the kinetic potential as 
\beq \label{K1_gen}
  \tilde K^{(1)}(M,\bar M,\cN+\bar \cN|L,\xi^{i}) = \tilde  K - \tfrac12(\cN_\kappa+\bar \cN_\kappa) \xi^\kappa\ .
\eeq
It is important to stress that in this expression one has to evaluate $\xi^\kappa$ as
function of $\cN_\kappa,M^I$ and $R,\xi^i$ by solving \eqref{N=dtildeK}. 

The explicit evaluation of  $N_k$ and $\tilde K^{(1)}$ is straightforward 
since $\tilde K$, given in \eqref{d=3_red_potential}, has such a simple form in this situation.
Evaluating $\partial_{\xi^\kappa} \tilde K$ one finds 
\beq
  \cN_\kappa =  R^{-1} \R f_{\kappa \Lambda} \xi^{\Lambda} - i\tilde \xi_\kappa  = f_{\kappa \Lambda} \zeta^{\Lambda} - i    \tilde b_\kappa\ ,
\eeq 
where we defined $\tilde b_\kappa = \tilde \xi_\kappa + \I f_{\kappa \Lambda} \zeta^{\Lambda}$.
Solving for $\xi^\kappa$ and inserting the result in \eqref{K1_gen} yields the 
expression
\beq \label{K1_exp}
  \tilde K^{(1)} = K(M,\bar M) + \log R - \tfrac12 R^{-1} \Delta_{ij}\, \xi^i \xi^j  + \tfrac12 R\, \Delta^{\kappa \lambda} \, \R \cN_\kappa \R \cN_\lambda -\, \Delta^\lambda_i \xi^i \, \R \cN_\lambda  \ ,
\eeq
where 
\beq \label{def-deltaapp}
  \Delta_{ij} = \R f_{ij} -  \R f_{\kappa i} \R f^{\kappa \lambda} \R f_{\lambda j}\ ,\qquad \Delta^{\kappa \lambda} =\R f^{\kappa \lambda}\ ,\qquad \Delta^\lambda_i = \R f^{\kappa \lambda}\, \R f_{\kappa i}\ .
\eeq
Using this kinetic potential in \eqref{kinetic_lin_gen}
one obtains the three-dimensional effective action for the chiral multiplets with complex scalars $(M^I,\cN_\kappa)$
and the vector multiplets $(\xi^i,A^i)$. 
The four-dimensional $\cN=1$ K\"ahler potential $K(M,\bar M)$ and gauge-kinetic coupling 
function $f_{\Lambda \Sigma}(M)$ are then determined comparing the F-theory 
kinetic potential with the general expression \eqref{K1_exp}.

\section{Compactifications with general three-forms on the Calabi-Yau fourfold} \label{allH3}

In order to proceed we need to dualize some of the vector multiplets $(L^\cA,A^\cA)$
into chiral multiplets. We will precisely pick those which correspond to 
divisors $D_\alpha$. We thus split 
\bea
   L^\cA = (L^\alpha|\xi^i,R)\ , &&\qquad  \quad \alpha = 1,\ldots,h^{1,1}(B_3)\\
   A^\cA = (A^\alpha|A^i,A^0)\ , &&\qquad \quad i = h^{1,1}(B_3)+1,\ldots,h^{1,1}(\hat {X}_4)-1\ . \nn
\eea
Our aim is to dualize the vector multiplets $(L^\alpha,A^\alpha)$
into chiral multiplets with two scalars $(L^\alpha,\rho_\alpha)$
which combine into complex coordinates $T_\alpha$.
We also split the set of complex scalars $N^I$ as 
\bea
    N^I = (\Nk,\Na)\ , &&\qquad  \quad \kappa =  1, \ldots, h^{2,1}(B_3)\ , \\
     &&\qquad  \quad a = h^{2,1}(B_3)+1 ,\ldots,h^{2,1}(\hat {X}_4)\ .\nn
\eea
This is again done to distinguish three-dimensional multiplets which descend to 
chiral or vector multiplets in four dimensions. More precisely, 
the $\Na$ will descend to complex scalars while the $\Nk$ become vectors in four 
space-time dimensions. 
We summarize this spectrum in table \ref{F-theory_spec_tab}.

The chiral multiplets and their corresponding kinetic potential 
are obtained by a Legendre transformation similar to 
the analysis of section \ref{scalar_vec_duality}. One thus defines 
the new complex coordinates 
\beq \label{Talpha_Leg}
   T_\alpha = \partial_{L^\alpha} \tilde K^{\rm M}+ i \rho_\alpha\ ,
\eeq
and evaluates the effective action starting with 
a new kinetic potential
\beq \label{kin_F-dual}
  \tilde K^{\rm F}(z,\bar z,N+\bar N,T+\bar T|\xi^i,R) = \tilde K^{\rm M} - \tfrac12 (T_\alpha + \bar T_\alpha) L^\alpha\ ,
\eeq
such that 
\beq
   \frac{\partial \tilde K^{\rm F}}{\partial T_\alpha }= - \frac12 L^\alpha \   \ , \qquad \frac{\partial \tilde K^{\rm F}}{\partial M} = \frac{\partial \tilde K^{\rm M}}{\partial M}\ , \qquad M \in (N^I,\xi^i,R,z)\ .
\eeq
Note that the right-hand sides of these expressions are evaluated by  first taking derivatives of $\tilde K^{\rm M}$ viewed 
as a function of $L^{\cA} =(L^\alpha,\xi^i,R)$ and $N^I$, and then use \eqref{Talpha_Leg} to express the 
result as a function of $T_\alpha,R,\xi^i,N^I$. One thus uses, for example, the identity
\bea
  \partial_{\xi^i} \big[\tilde K^{\rm M}(T,N|\xi) \big] &=& \partial_{\xi^i} \big[\tilde K^{\rm M}(N|\xi,L(T,\xi,N)) \big] \\
  &=& \partial_{\xi^i} \big[\tilde K^{\rm M}(N|\xi,L) \big] + \partial_{L^\alpha} \big[\tilde K^{\rm M}(N|\xi,L) \big] \frac{\partial L^\alpha}{\partial \xi^i}\ , \nn
\eea 
where we have suppressed the dependence on $z$ and $R$ to make the expressions more readable.
Note that by 
differentiating \eqref{Talpha_Leg} one also finds
\beq
  \frac{\partial L^\alpha}{ \partial T_\beta} =  \tilde K^{{\rm M}\ L^\alpha L^\beta} \ ,\qquad   
  \frac{\partial L^\alpha}{\partial M} = - \tilde K^{{\rm M}\ L^\alpha L^\beta} \partial_M \tilde K^{\rm M}_{L^\beta}\ , \qquad M \in (N^I,\xi^i,R,z)\ .
\eeq

However, since the kinetic 
potential $\tilde K^{\rm M}$  before dualization 
is in general very complicated the resulting expression for the 
dual theory turns out to be rather involved. Let us make some 
observations first. Note that $\xi^i$ only appears linearly 
and quadratically in the general expression \eqref{K1_exp}
since $\xi^i$ arises as the fourth component of $D=4$ 
vectors and we only included the standard Yang-Mills terms 
for these fields. Hence, we expand $\tilde K^{\rm F}$, given 
in \eqref{kin_F-dual}, to quadratic order around $\xi^i = 0$, $\Nk=0$
and note that it should take the form 
\beq \label{K1_F-theory}
  \tilde K^{(1)} = \tilde K^{\rm F}\big|_{*} - \tfrac12 R^{-1} \ \Delta_{ij}\, \xi^i \, \xi^j  + \tfrac12 R\ \Delta^{\kappa \lambda}\, \R \Nk \, \R \Nl - \Delta_{i}^\kappa\, \xi^i \, \R \Nk\ .
\eeq
where the $*$ indicates that the expression has to be evaluated at $\xi^i=0,\R \Nk=0$.
Here we have abbreviated 
\beq
  \begin{array}{rclcl}
  \Delta_{ij} &=&  - R\, \partial_{\xi^i} \partial_{\xi^j}\tilde K^{\rm F}\big|_* & \stackrel{!}{=} & \R f_{ij} - \R f_{i \kappa}\ \R f^{\kappa \lambda}\ \R f_{\lambda j} \ , \\[.2cm]
  \Delta_{i}^\kappa &=& - \partial_{\xi^i} \partial_{\R \Nk}\tilde K^{\rm F}\big|_* & \stackrel{!}{=}& \R f_{i \kappa}\ \R f^{\kappa \lambda} \ , \\[.2cm]
  \Delta^{\kappa \lambda} &=&  R^{-1}\, \partial_{\R \Nk} \partial_{\R \Nl}\tilde K^{\rm F}\big|_*\quad & \stackrel{!}{=}&\R f^{\kappa \lambda} \ .
  \end{array}
\eeq
Here the expressions after the second equal signs in each line are the expected expressions 
obtained by comparison with \eqref{K1_exp}.
It will be the kinetic potential \eqref{K1_F-theory}, which one compares to the expression \eqref{K1_exp}
to extract the $\cN=1,D=4$ characteristic data of F-theory on a Calabi-Yau fourfold.
Since there is no linear term in \eqref{K1_exp} we thus expect to find 
\bea
  \partial_{\xi^j}\tilde K^{\rm F}\big|_* = \partial_{\xi^j}\tilde K^{\rm M}\big|_* = 0 \ ,\qquad \partial_{\Nk}\tilde K^{\rm F}\big|_* = \partial_{\Nk}\tilde K^{\rm M}\big|_* = 0\ ,
\eea
in a set-up which consistently lifts to a four-dimensional $\cN=1$ F-theory compactification.
Similarly one 
evaluates the second derivatives and translates the derivatives of $\tilde K^{\rm F}$ into derivatives of $\tilde K^{\rm M}$.


\end{document}